\documentclass[aps,preprint,nofootinbib]{revtex4}%
\usepackage{amsmath}
\usepackage{graphicx}
\usepackage{amsfonts}
\usepackage{amssymb}%
\providecommand{\U}[1]{\protect\rule{.1in}{.1in}}
\providecommand{\U}[1]{\protect\rule{.1in}{.1in}}
\providecommand{\U}[1]{\protect\rule{.1in}{.1in}}
\providecommand{\U}[1]{\protect\rule{.1in}{.1in}}
\providecommand{\U}[1]{\protect\rule{.1in}{.1in}}
\providecommand{\U}[1]{\protect\rule{.1in}{.1in}}
\providecommand{\U}[1]{\protect\rule{.1in}{.1in}}
\providecommand{\U}[1]{\protect\rule{.1in}{.1in}}
\providecommand{\U}[1]{\protect\rule{.1in}{.1in}}
\providecommand{\U}[1]{\protect\rule{.1in}{.1in}}
\providecommand{\U}[1]{\protect\rule{.1in}{.1in}}

\begin{document}
\preprint{ }
\title{}
\author{}
\maketitle

\begin{center}
{\Large Geodesically Complete Analytic Solutions for a Cyclic Universe
}\footnote{Work partially supported by the US Department of Energy, grant
number DE-FG03-84ER40168.}

{\vskip0.5cm}

\textbf{Itzhak Bars}$^{\ast\#}$\textbf{\ , Shih-Hung Chen}$^{\dagger\#}%
$\textbf{\ and Neil Turok}$^{\#}$

{\vskip0.5cm}

$^{\ast}$\textsl{Department of Physics and Astronomy}

\textsl{University of Southern California,\ Los Angeles, CA 90089-2535 USA}

\vspace{0.5cm}

$^{\dagger}$\textsl{Department of Physics and School of Earth and Space
Exploration}

\textsl{Arizona State University, Tempe, AZ 85287-1404 USA}

\vspace{0.5cm}

$^{\#}$\textsl{Perimeter Institute for Theoretical Physics}

\textsl{Waterloo, ON N2L 2Y5, Canada}

\vspace{0.5cm}

{\vskip0.4cm} \textbf{Abstract}
\end{center}

We present analytic solutions to a class of cosmological models described by a
canonical scalar field minimally coupled to gravity and experiencing self
interactions through a hyperbolic potential. Using models and methods inspired
by 2T-physics, we show how analytic solutions can be obtained in
flat/open/closed Friedmann-Robertson-Walker universes. Among the analytic
solutions, there are many interesting geodesically complete cyclic solutions
in which the universe bounces at either zero or finite sizes. When geodesic
completeness is imposed, it restricts models and their parameters to a certain
parameter subspace, including some quantization conditions on initial
conditions in the case of zero-size bounces, but no conditions on initial
conditions for the case of finite-size bounces. We will explain the
theoretical origin of our model from the point of view of 2T-gravity as well
as from the point of view of the colliding branes scenario in the context of
M-theory. We will indicate how to associate solutions of the quantum
Wheeler-deWitt equation with our classical analytic solutions, mention some
physical aspects of the cyclic solutions, and outline future directions.



\newpage
\tableofcontents

\newpage

\section{Introduction}

\label{intro}

In this paper we will analytically express a cyclic universe using exact
solutions in a scalar-tensor theory with a scalar field $\sigma\left(  x^{\mu
}\right)  $ minimally coupled to gravity.

The full action of our theory is%
\begin{equation}
S=\int d^{4}x\sqrt{-g}\left\{  \frac{1}{2\kappa^{2}}R\left(  g\right)
-\frac{1}{2}g^{\mu\nu}\partial_{\mu}\sigma\partial_{\nu}\sigma-V\left(
\sigma\right)  \right\}  , \label{theory}%
\end{equation}
where the potential is
\begin{equation}
V\left(  \sigma\right)  =\left(  \frac{\sqrt{6}}{\kappa}\right)  ^{4}\left[
b\cosh^{4}\left(  \frac{\kappa\sigma}{\sqrt{6}}\right)  +c\sinh^{4}\left(
\frac{\kappa\sigma}{\sqrt{6}}\right)  \right]  . \label{potential}%
\end{equation}
Here $b$ and $c$ are dimensionless free parameters of the potential, and
$\kappa^{-1}$ is the reduced Planck mass $\kappa^{-1}=\sqrt{\frac{\hbar
c}{8\pi G}}=2.43\times10^{18}\frac{GeV}{c^{2}}$. A plot of the potential
energy $V\left(  \sigma\right)  $ for various signs and magnitudes of $b,c,$
consistent with stability $(b+c)>0,$ show that the profile of this potential
is similar to those often used in the study of cosmology. This potential was
chosen because we can solve the equations exactly, thus enabling us to perform
the type of analysis presented in this paper. We assume that the general
features discussed here go beyond the special choice of potential\footnote{In
fact, this is not the only potential for which we are able to give a full
analysis with the complete set of analytic solutions \cite{bcst2}. In the near
future we will present a similar discussion for the potentials $V_{1}\left(
\sigma\right)  =\left(  \frac{\sqrt{6}}{\kappa}\right)  ^{4}\left(
be^{-2\kappa\sigma/\sqrt{6}}+ce^{-4\kappa\sigma/\sqrt{6}}\right)  $ and
$V_{2}\left(  \sigma\right)  =\left(  \frac{\sqrt{6}}{\kappa}\right)
^{4}be^{2p\kappa\sigma/\sqrt{6}}$, where $b,c,p$ are dimensionless real
parameters. The profile of $V_{1}\left(  \sigma\right)  ,$ with $c>0$ and
$b<0,$ is similar to the profile of the potential used initially in the cyclic
cosmology model in \cite{cyclicST}. \label{others}}.

The model of Eqs.(\ref{theory},\ref{potential}) was initially inspired by 2T
physics \cite{2TPhaseSpace}\cite{2Tgravity}\cite{2TgravityGeometry} as
described in \cite{inflationBC} and section (\ref{2Torigin}) below. The same
model fits also in the worldbrane scenario \cite{RandallSundrum}, as inspired
by D-branes in M-theory \cite{HoravaWitten}. The ideas of a cyclic universe
\cite{cyclicST} modeled in Ref.\cite{branesMT} can be adapted to reproduce the
same potential $V\left(  \sigma\right)  $, thus describing a universe that
consists of two 3+1 dimensional orientifolds that periodically collide with
each other by oscillating in an extra fifth dimension. It is quite interesting
that this connection emerged between 2T-gravity and M-theory. In subsections
(\ref{2Torigin}) and (\ref{Neil}) we will comment on the different origins
that converged on this model.

In a previous cosmological application of this model \cite{inflationBC}
$V\left(  \sigma\right)  $ was an energy density of the order of the grand
unification scale $\left(  m_{GUT}\right)  ^{4}$. In that case, $b^{1/4}%
\sqrt{6}\kappa^{-1}$ or $c^{1/4}\sqrt{6}\kappa^{-1}$ were of order
$m_{GUT}\sim10^{16}\frac{GeV}{c^{2}},$ thus leading to dimensionless values
for the parameters $b$ or $c$ in the order of $10^{-12}$. Exact solutions have
a way of finding applications in various fields. For physical applications of
our solutions, including cyclic cosmology or other future cases, the value of
the parameters $b,c,$ should be chosen appropriately depending on the application.

The complete set of analytic solutions for this model, in a homogeneous,
spatially flat, isotropic Friedmann-Robertson-Walker (FRW) universe, were
obtained in our earlier paper \cite{inflationBC}, and some of their
perturbations were studied in \cite{inflationBC}\cite{perturbationCD}. In the
current paper we will emphasize a subset of these solutions that are
geodesically complete and describe a universe smoothly evolving through big
bang or big crunch singularities at which the universe shrinks to \textit{zero
size}, but then it continues to perform periodic expansions and contractions
that describe a cyclic universe, all without violating unitarity or the null
energy condition in a flat universe. We will also include the effect of
spacial curvature for the FRW universes ($k=0,\pm1$) in our new exact
solutions, and we will exhibit cyclic solutions with \textit{finite size
bounces} as well.

Perturbations such as radiation is easily included in the exact solutions,
while anisotropy can be discussed with analytic approximations; but those
aspects, as well as the quantum treatment through the Wheeler-deWitt equation,
which require more detailed discussions, will appear in a separate paper
\cite{bcst1}.

The complete set of homogeneous, isotropic classical solutions
presented in \cite{inflationBC} show that, the \textit{generic}
solutions for the field $\sigma\left(  \tau\right)  $ and the scale
factor $a\left(  \tau\right)  $ describe a geodesically
\textit{incomplete} geometry. The geodesic incompleteness can be
exhibited in terms of conformal time $\tau$ as defined by the line
element $ds^{2}=a^{2}\left(  \tau\right)  \left(  -d\tau
^{2}+ds_{3}^{2}\right)  ,$ where $ds_{3}^{2}$ is the line element of
the 3-dimensional space. As an illustration consider the spatially
flat case $ds_{3}^{2}=d\vec{x}\cdot d\vec{x}.$ The geodesic
$x^{\mu}\left(  \tau\right) $ of a massive particle in this flat
geometry is described by its velocity
\begin{equation}
\frac{d\vec{x}\left(  \tau\right)  }{d\tau}=\frac{\vec{p}}{\sqrt{\vec{p}%
^{2}+m^{2}a^{2}\left(  \tau\right)  }},
\end{equation}
where $\vec{p}$ is the conserved 3-momentum. In terms of this
conformal time $\tau$, the complete set of solutions in
\cite{inflationBC} show that, for the generic solution, the scale
factor $a\left(  \tau\right)  $ starts out at zero size at some time
$a\left(  \tau_{1}\right)  =0$ and grows to maximum size $a\left(
\tau_{2}\right)  =a_{\max}$ in a finite amount of conformal time
$\left(  \tau_{2}-\tau_{1}\right)  =$finite. It turns out that
$a_{\max}$ is infinite in the case of $b\geq0$ or finite in the case
of $b<0$. Furthermore $a\left(  \tau\right)  $ has this same
behavior in an infinite number of different disconnected intervals
of conformal time $\tau.$ Each such separate interval describes a
universe that starts out with a big bang and expands to maximum
size. Moreover there are other disconnected intervals in which the
universe contracts from maximum size to zero size. Evidently such
generic solutions of the Friedmann equations are geodesically
incomplete.

If expressed in terms of cosmic time $t$ defined by the line element
$ds^{2}=\left(  -dt^{2}+a^{2}\left(  t\right)  ds_{3}^{2}\right)  ,$
the geodesic equation reads
\begin{equation}
\frac{d\vec{x}\left(  t\right)  }{dt}=\frac{\vec{p}}{a\left(
t\right)
\sqrt{\vec{p}^{2}+m^{2}a^{2}\left(  t\right)  }}%
\end{equation}
where $a\left(  t\right)  $ is expressed in terms of cosmic time
$t$. The relation to conformal time is $dt=a\left(  \tau\right)
d\tau$ or $t\left( \tau\right)  =\int_{\tau_{1}}^{\tau}a\left(
\tau\right)  d\tau,$ where $t\left(  \tau_{1}\right)  =0$ defines
the big bang at $a\left(  t\left( \tau_{1}\right)  \right)  =0$.
Hence $t\left(  \tau\right)  $ is given by the area under the curve
in a plot of $a\left(  \tau\right)  $ versus $\tau,$ for some
interval $\tau_{1}\leq\tau\leq\tau_{2},$ starting with the big bang.
An example of a geodesically incomplete curve $a\left(  \tau\right)
$ is Fig.1 in \cite{inflationBC} while examples of geodesically
complete ones are given in many figures in the current paper. Since
$a\left(  \tau\right)  $ in the generic solution is given in
disconnected $\tau$ intervals, the cosmic time $t\left(  \tau\right)
$ cannot be defined for negative values before the big bang. Hence
the geodesic equation above is artificially stopped at the big bang
at the finite value of time $t\left(  \tau_{1}\right)  =0$. This is
one of the signs of geodesic incompleteness of this geometry. In
addition, when the area under the curve is finite (the solutions for
$b<0$), the total cosmic time $t\left(  \tau_{2}\right)  $ is also
finite, and geodesics are again artificially stopped at a finite
value of cosmic time $t\left(  \tau _{2}\right)  ,$ providing
another sign of geodesic incompleteness. In this way, each interval
that is geodesically incomplete in conformal time appears again as
geodesically incomplete in cosmic time. This type of geometry
bounded by singularities, and classical solutions in them, occur
often in General Relativity and are commonly used in its
applications, as in our own paper \cite{inflationBC}. But in view of
the geodesic incompleteness of the generic solutions of the
Friedmann equations displayed in the conformal frame, it feels that
this must be an incomplete story.

We think that a more satisfactory approach, especially for cosmology, is to
find those solutions that describe a geodesically complete geometry. This type
of solution is what we will describe in the current paper. It turns out that
among the classical solutions presented in \cite{inflationBC} there are some
unique solutions that are geodesically complete in the \textit{Einstein
frame}. In this solution the patches of conformal time in which $a\left(
\tau\right)  $ is real in the Einstein frame\footnote{All solutions, including
those that are geodesically incomplete in the Einstein frame, are geodesically
complete in other frames. Indeed, in the $\gamma$-gauge that we will discuss
in section (\ref{gamma}), all solutions are geodesically complete. However, as
viewed from the point of view of the Einstein frame, $a\left(  \tau\right)  $
for such solutions becomes imaginary in the patches that complete the
geodesics in the $\gamma$-frame, and hence the physical meaning in the
Einstein frame becomes obscure. We intend to study this phenomenon in a future
paper, but for the current paper we concentrate on only the geodesically
complete solutions in which $a\left(  \tau\right)  $ is real for all $\tau.$}
are smoothly connected from $\tau=-\infty$ to $\tau=\infty.$ Then the universe
sails smoothly through singularities, while geodesics of both massless and
massive particles smoothly continue through singularities to the next cycle of
the cyclic universe.

The requirements for such solutions depend on whether the bounce is at zero
size $a\left(  \tau_{bang}\right)  =0$ or finite size $a\left(  \tau
_{bang}\right)  >0$. In the case of zero size bounce, that occurs when the
spacial curvature is zero, $k=0$, initial conditions of the two fields need to
be synchronized and periods of oscillation need to be relatively quantized, as
we will described in detail. These requirements result in some quantization
conditions among the available parameters consisting of integration constants
of the differential equations for $\sigma\left(  \tau\right)  ,a\left(
\tau\right)  $ as well as the parameters $b,c$ in the potential energy
$V\left(  \sigma\right)  ,$ and also radiation and anisotropy parameters when
they are included. Because of these requirements these geodesically complete
zero-size bounce solutions are associated with a countable set of boundary
conditions (but still an infinite set, in the sense described in section
(\ref{geodC})). In the case of finite-size bounce none of these constraints
occur on boundary conditions, but in this case there is spacial curvature,
$k/r_{0}^{2}\neq0,$ which needs to be large enough to compete with the
potential energy. In this case, as long as the parameters that define the
model are within a certain continuous region, the generic solution is the
finite-size bounce solution without any further requirements on boundary conditions.

When perturbations, such as radiation and anisotropy are included, or when
quantum effects in the form of the Wheeler-deWitt equation are taken into
account, there still are geodesically complete solutions that have a similar
character to what we will present in this paper. They still form a countable
set for zero size bounces, while they are the generic solutions for the finite
bounces. Either way, the distinguishing character of geodesic completeness has
an appeal that seems important for physical applications, as we will discuss
in a future paper \cite{bcst1}.

\subsection{2T-physics origin \label{2Torigin}}

We would like to briefly summarize here the main points of how the model in
Eq.(\ref{theory}) relates to 2T-gravity \cite{2Tgravity}%
\cite{2TgravityGeometry} in 4-space and 2-time dimensions as the conformal
shadow in 3-space and 1-time. More generally, according to 2T-physics, a
theory in 1T-physics in $\left(  d-1\right)  +1$ dimensions is one of the many
shadows that comes from $d+2$ dimensions \cite{2TPhaseSpace}. A useful shadow
that appeals to the intuition of physicists accustomed to relativistic field
theory is the one called the \textit{conformal shadow}. In the conformal
shadow there is a local scale symmetry (Weyl symmetry). The original
2T-gravity in 4+2 dimensions does not have a Weyl symmetry in 4+2 dimensions,
instead this crucial gauge symmetry in 3+1 dimensions is dictated in the
conformal shadow as a remnant of general coordinate transformations in the
extra 1+1 dimensions \cite{2TgravityGeometry}. Other less familiar shadows
provide other descriptions of the physics that are related by duality
transformations to the conformal shadow, and often they can provide hidden
physical information that is systematically missed in the conventional
formulation of 1T-physics \cite{2TPhaseSpace}\cite{emergentfieldth1}%
\cite{emergentfieldth2}.

Besides this duality aspect, 2T-physics may also provide additional
constraints on the interactions of fields in 1T-physics even within the
conformal shadow. The constraints in theories of interest, in the conformal
shadow, are mainly on scalar fields and their interactions. These constraints
have been determined generally in \cite{2Tsugra} in the presence of gravity or
supergravity. Most of the emergent constraints can be rephrased as being
consequences of various symmetries in 1T-field theory, but not all of them.
Some of those additional constraints are not motivated by fundamental
principles in 1T-physics, as discussed in \cite{2Tsugra}, so they can be taken
as signatures of 2T-physics. Here we will deal only with the simplest version
of scalar fields that obey the constraints. This is the case when \textit{all
scalar fields} are conformally coupled to gravity in the 1T version. This is a
familiar form in 1T field theory, but 1T-gravity does not require that
\textit{all} the scalars be conformal scalars; by contrast 2T-gravity has this
as an outcome for the conformal shadow in one of its allowed versions (more
general form of constrained scalars in \cite{2Tsugra}).

Moreover, in the conformal shadow there is no Einstein-Hilbert term in the
action, so there is no dimensionful gravitational constant that would break
the local Weyl symmetry explicitly. Instead, the Einstein-Hilbert term emerges
from gauge fixing the Weyl symmetry within the conformal shadow (see below).
This is a mechanism called \textquotedblleft compensating
fields\textquotedblright\ which is familiar in conventional field theory. Such
structures of 2T-physics are compatible with the construction of satisfactory
models of a complete theory of Nature directly in 4+2 dimensions, including
the standard model \cite{2tstandardM}, its generalizations with supersymmetry
\cite{susy2tN1}\cite{susy2tN2N4} or grand unification, gravity
\cite{2Tgravity}\cite{2TgravityGeometry}, supergravity \cite{2Tsugra}, all of
which lead to applications in LHC physics and cosmology \cite{inflationBC}.

The derivation of the \textit{conformal shadow} in 3+1 dimensions from the 4+2
dimensional theory is described in detail in \cite{2Tgravity}%
\cite{2TgravityGeometry}. For the purpose of the current paper it is possible
to skip this detail and start out directly in 3+1 dimensions by requiring a
local scale symmetry (Weyl symmetry). Then, the ordinary looking model in
Eqs.(\ref{theory},\ref{potential}) can be presented as a gauge fixed version
of the following gauge invariant field theory in 3+1 dimensions, which is a
conformal\textit{ }shadow that contains one scalar field $s\left(  x\right)  $
in addition to a dilaton $\phi\left(  x\right)  $, both conformally coupled to
gravity as follows
\begin{equation}
S=\int d^{4}x\sqrt{-g}\left(  \frac{1}{2}g^{\mu\nu}\partial_{\mu}\phi
\partial_{\nu}\phi-\frac{1}{2}g^{\mu\nu}\partial_{\mu}s\partial_{\nu}%
s+\frac{1}{12}\left(  \phi^{2}-s^{2}\right)  R\left(  g\right)  -\phi
^{4}f\left(  \frac{s}{\phi}\right)  \right)  . \label{initialS}%
\end{equation}
The field $\phi\left(  x\right)  $ has the wrong sign in the kinetic term, so
it is a ghost (negative norm\footnote{There are models of cosmology based on
the notion of \textquotedblleft quintom matter\textquotedblright%
\ \cite{quintom} which also introduce a negative norm ghost field. We should
emphasize that those models have actual ghosts and therefore are non-unitary
and fundamentally flawed. Despite some similarity, our model is fundamentally
different because of the Weyl symmetry that eliminates the ghosts, thus having
fewer degrees of freedom. Our action, our solutions which do not violate the
null energy conditions, and the discussion of the physics are also
different.}). This sign of the kinetic term is required by the Weyl symmetry
if the sign in front of the curvature term $\frac{1}{12}\phi^{2}R\left(
g\right)  $ is positive. However, due to the local scale symmetry the ghost
can be gauge fixed away, so this theory is unitary. The gauge symmetry is
preserved for any potential of the form $\phi^{4}f\left(  \frac{s}{\phi
}\right)  $ where $f\left(  z\right)  $ is an arbitrary function of its
argument $z=\frac{s}{\phi}$. In this action there is no Einstein-Hilbert term
with a dimensionful gravitational constant, but instead, the factor $\left(
\phi^{2}-s^{2}\right)  ^{-1}$ plays the role of a spacetime dependent
effective \textquotedblleft gravitational parameter\textquotedblright.

\subsection{Braneworld origin \label{Neil}}

A cyclic model, inspired by D-branes in M-theory \cite{HoravaWitten}, was
developed in \cite{cyclicST} where it was discussed for a very different
potential than Eq.(\ref{potential}). However, it is possible to recover
precisely the current model of Eqs.(\ref{theory},\ref{potential}) in the
colliding world brane scenario as follows. One should compare Eq.(27) in
Ref.\cite{branesMT} to the model in Eq.(\ref{initialS}) before gauge fixing
the Weyl symmetry. Both models have a Weyl symmetry that is a remnant of
general coordinate transformations in extra dimensions (although the extra
dimensions in the two cases do not have identical signatures). One should
compare our fields here $s\left(  x\right)  ,\phi\left(  x\right)  $ to the
fields $\psi^{\pm}\left(  x\right)  $ in Ref.\cite{branesMT}, since they are
both conformally coupled scalars and have precisely the same kinetic energy
terms. Furthermore, the potential $V\left(  \sigma\right)  $ of
Eq.(\ref{potential}) is also recovered, if one replaces the unknown terms in
Eq.(27) of Ref.\cite{branesMT}, $2W_{CFT}\left[  g^{+}\right]  -2W_{CFT}%
\left[  g^{-}\right]  +S_{m}[g^{+}]+S_{m}[g^{-}],$ by just a cosmological term
on each brane. Namely, replacing the unknown expression by, $b_{+}\sqrt{g^{+}%
}+b_{-}\sqrt{g^{-}},$ where $b_{\pm}$ are constants, and using their
definition of $g^{\pm}$, gives $\sum_{\pm}b_{\pm}\sqrt{g^{\pm}}=\sum_{\pm
}b_{\pm}\sqrt{-g}\left(  \psi^{\pm}\right)  ^{4}.$ This is indeed the
potential $b\phi^{4}+cs^{4}$ in Eq.(\ref{initialS}), which in turn leads to
the potential $V\left(  \sigma\right)  $ after the Weyl gauge symmetry is
fixed to obtain the Einstein frame as described in \cite{inflationBC} and below.

\subsection{Fixing the Weyl symmetry}

The Weyl symmetry can be gauge fixed in several forms. In the Einstein gauge
denoted by a label $E,$ such as $\phi_{E},s_{E},g_{E}^{\mu\nu}$, the gauge is
fixed such that the curvature term $\frac{1}{12}\left(  \phi^{2}-s^{2}\right)
R\left(  g\right)  $ becomes precisely the Einstein-Hilbert term $\frac
{1}{2\kappa^{2}}R\left(  g_{E}\right)  ,$ so that in the Einstein gauge we
have%
\begin{equation}
\frac{1}{12}\left(  \phi_{E}^{2}-s_{E}^{2}\right)  =\frac{1}{2\kappa^{2}}.
\label{Egauge}%
\end{equation}
In this gauge it is convenient to parametrize $\phi_{E},s_{E}$ in terms of a
single scalar field $\sigma$%
\begin{equation}
\phi_{E}\left(  x\right)  =\pm\frac{\sqrt{6}}{\kappa}\cosh\left(  \frac
{\kappa\sigma\left(  x\right)  }{\sqrt{6}}\right)  ,\;s_{E}\left(  x\right)
=\frac{\sqrt{6}}{\kappa}\sinh\left(  \frac{\kappa\sigma\left(  x\right)
}{\sqrt{6}}\right)  .
\end{equation}
Then the gauge fixed form of the action in Eq.(\ref{initialS}) takes precisely
the form of Eq.(\ref{theory}), where the potential $V\left(  \sigma\right)  $
is arbitrary as long as the function $f\left(  z\right)  $ is arbitrary.

The Friedmann-Robertson-Walker metric (FRW) in this gauge takes the form%
\begin{align}
ds_{E}^{2}  &  =-dt^{2}+a_{E}^{2}\left(  t\right)  ds_{3}^{2}=a^{2}\left(
\tau\right)  \left(  -d\tau^{2}+ds_{3}^{2}\right)  ,\label{Emetric}\\
ds_{3}^{2}  &  =\frac{dr^{2}}{1-kr^{2}/r_{0}^{2}}+r^{2}\left(  d\theta
^{2}+\sin^{2}\theta d\phi^{2}\right)  ;\;k=0,\pm1.
\end{align}
where $ds_{3}^{2}$ is the metric of the 3-dimensional space, $\tau$ is the
conformal time and $a\left(  \tau\right)  \equiv a_{E}\left(  t\left(
\tau\right)  \right)  $ is the cosmological scale whose dynamics we wish to
study in this paper. The relation between ordinary co-moving time $t$ and the
conformal time is\footnote{In this paper the overdot denotes derivative with
respect to conformal time $\dot{a}\left(  \tau\right)  \equiv da/d\tau$ and
$\ddot{a}\left(  \tau\right)  =d^{2}a/\left(  d\tau\right)  ^{2}.$ The
derivative with respect to comoving time $t$ can be rewritten by using the
chain rule as $\frac{d}{dt}=\frac{1}{a\left(  \tau\right)  }\frac{d}{d\tau}.$
For example, the Hubble parameter $H\equiv\frac{1}{a_{E}\left(  t\right)
}\frac{da_{E}}{dt}$ and its derivative $\frac{dH}{dt}+H^{2}=\frac{1}{a_{E}%
}\frac{d^{2}a_{E}}{\left(  dt\right)  ^{2}},$ are expressed as
\begin{equation}
H=\frac{\dot{a}\left(  \tau\right)  }{a^{2}\left(  \tau\right)  },\;\frac
{dH}{dt}+H^{2}=\frac{\overset{\cdot\cdot}{a}}{a^{3}}-\frac{\dot{a}^{2}}{a^{4}%
}.
\end{equation}
\label{hubble}}
\begin{equation}
dt=a\left(  \tau\right)  d\tau,\text{ or }t\left(  \tau\right)  =\int
_{0}^{\tau}a\left(  \tau^{\prime}\right)  d\tau^{\prime}.
\end{equation}
The scalar curvature of the metric in Eq.(\ref{Emetric}) is given by
\begin{equation}
R\left(  g_{E}\right)  =\frac{6}{a^{2}}(\frac{\ddot{a}}{a}+\frac{k}{r_{0}^{2}%
}),
\end{equation}
where $r_{0}$ is a constant radius that sets the scale of the curvature of
3-space\footnote{The parameter $r_{0}$ sets the scale for the curvature. If
normalized to today's curvature, with $r_{0}$ representing todays size of the
universe, then $K=k/r_{0}^{2}$ is extremely small even when $k\neq0$. In that
case we can completely forget the effect of spacial curvature. However, there
are cosmological models that play with the curvature parameter as applied in
the early stages of the universe. In that case $r_{0}$ may be within a few
factors of $10$ of the Planck scale, in which case the curvature is enormous.
In order not to miss on possible interesting solutions we will not pre-judge
the size of this term and investigate the solutions that emerge. Then in
various physical applications we may or may not neglect the parameter
$K=k/r_{0}^{2}.$} when the dimensionless scale factor is $a=1$.

Thus, in this gauge, for the purpose of homogeneous solutions of the equations
of motion, the dynamical variables are $a\left(  \tau\right)  $ and
$\sigma\left(  \tau\right)  $ which interact with each other as prescribed by
the action (\ref{theory}). Their equations of motion reduce to the Friedmann
equations \cite{Friedmann} as follows$^{\ref{hubble}}$
\begin{gather}
\frac{\dot{a}^{2}}{a^{4}}=\frac{\kappa^{2}}{3}\left[  \frac{\dot{\sigma}^{2}%
}{2a^{2}}+V\left(  \sigma\right)  \right]  -\frac{k}{r_{0}^{2}a^{2}%
},\label{00}\\
\frac{\ddot{a}}{a^{3}}-\frac{\dot{a}^{2}}{a^{4}}=-\frac{\kappa^{2}}{3}\left[
\frac{\dot{\sigma}^{2}}{a^{2}}-V\left(  \sigma\right)  \right]  ,\label{11}\\
\frac{\ddot{\sigma}}{a^{2}}+2\frac{\dot{a}\dot{\sigma}}{a^{3}}+V^{\prime
}\left(  \sigma\right)  =0, \label{eom sigma}%
\end{gather}
We had previously found all the exact solutions of these equations for the
potential $V\left(  \sigma\right)  $ given in Eq.(\ref{potential}) and a flat
universe $k=0$. These were tabulated in \cite{inflationBC}. In this paper we
will emphasize the subset of those solutions that are geodesically complete
and in addition we will generalize them by including non-zero spacial
curvature $k=\pm1.$ Further generalizations including radiation and
anisotropic metrics will be given in \cite{bcst1}. To explain what we mean by
a geodesically complete solution we need the following discussion.

\subsection{Geodesic completeness \label{geodC}}

Note that the Einstein gauge in Eq.(\ref{Egauge}) can be chosen only in
patches of spacetime $x^{\mu}$ when the \textit{gauge invariant} quantity
$\left[  1-s^{2}\left(  x^{\mu}\right)  /\phi^{2}\left(  x^{\mu}\right)
\right]  $ is positive. This quantity may be expressed in the Einstein gauge
(i.e. when it is positive only) as $\left(  1-s_{E}^{2}/\phi_{E}^{2}\right)
=\left(  \cosh(\kappa\sigma/\sqrt{6})\right)  ^{-2}.$ We must note that this
gauge invariant quantity could vanish at various times $\tau$. We did in fact
find that it does vanish at various values of $\tau$ in \textit{generic}
analytic solutions for $\sigma\left(  \tau\right)  ,a\left(  \tau\right)  $
given in \cite{inflationBC}. However, when $\left(  1-s_{E}^{2}/\phi_{E}%
^{2}\right)  $ vanishes $\phi_{E}$ diverges so as to maintain the gauge choice
for the gauge dependent quantity $\left(  \phi^{2}-s^{2}\right)  $ as given in
Eq.(\ref{Egauge}). But the Einstein gauge was fixed under the assumption that
$\left(  \phi^{2}-s^{2}\right)  $ was positive; if it can vanish can it also
change sign? This is the question that initially motivated two of us
\cite{inflationBC} to study this model in the $\phi,s$ version, rather than
the $a,\sigma$ version. Will the dynamics require the \textit{gauge invariant}
quantity $\left(  1-s^{2}/\phi^{2}\right)  $ to change sign in some patches of
spacetime, thus creating patches with antigravity? If yes, what would that
mean cosmologically for the universe we live in?

In our previous study in \cite{inflationBC} our exact solutions for $\phi,s$
showed that generically the dynamics does require the \textit{gauge invariant}
$\left(  1-s^{2}/\phi^{2}\right)  $ to change sign. However, the point at
which $\left(  1-s^{2}/\phi^{2}\right)  $ vanishes corresponds to a big bang
or a big crunch singularity where the scale factor in the Einstein gauge
vanishes $a^{2}\left(  \tau\right)  =a_{E}\left(  t\left(  \tau\right)
\right)  =0$ (recall $a\left(  \tau\right)  $ is \textit{gauge dependent}), so
the physical interpretation for our own universe may be stopped exactly where
$\left(  1-s^{2}/\phi^{2}\right)  $ vanishes, and therefore the solution could
be stopped artificially at that moment in conformal time $\tau.$ This is
geodesically incomplete, as well as gauge dependent from the point of view of
$a\left(  \tau\right)  $ as defined in the Einstein frame. But nevertheless,
if one insists that the theory is defined only in the Einstein frame, one
could accept a geodesically incomplete patch for a physical interpretation in
the usual interpretation of gravity. This type of geodesically incomplete
solution, which is very common in applications of gravity, was used in the
application to an inflating universe we discussed in our previous paper
\cite{inflationBC}.

In the current paper we take the point of view that geodesically complete
solutions are more satisfactory. To discover and better understand the
solution, we examine the factor $\left(  \phi^{2}-s^{2}\right)  $ that
multiplies $R\left(  g\right)  $ in the action. To overcome the gauge
dependent description, of the Einstein or other frames, we focus on the gauge
invariant quantity $\left(  1-s^{2}/\phi^{2}\right)  $. The point at which it
vanishes corresponds to the big bang or big crunch. When it is positive we can
choose the Einstein gauge to describe ordinary gravity, but in patches when it
is negative there is antigravity. Only the geodesically complete solutions has
no antigravity by having $\left(  1-s^{2}/\phi^{2}\right)  \geq0$ as a
function of $\tau$. Of course, quantum corrections near singularities may
smooth out the behavior of solutions. Notwithstanding the cloudiness of our
understanding of quantum gravity at this time, it still seems to us attractive
to identify the geodesically complete geometries and solutions in the
applications to cosmology, expecting that this feature survives the quantum
effects, as it seems to be the case at the level of the Wheeler-deWitt
equation \cite{bcst1}. Hence we will identify the circumstances in which there
are geodesically complete solutions in which the quantity $\left(
1-s^{2}/\phi^{2}\right)  $ never changes sign\footnote{I. Bars thanks Paul
Steinhart for stimulating discussions that led us to focus on this question.}.
We found that this is indeed possible, and we explicitly obtained those unique
geodesically complete solutions that are presented in the present paper and in
\cite{bcst1}.

As we will see in the explicit solutions given below, it turns out geodesic
completeness, for the \textit{bounce at zero size}, requires two ingredients.
First, the parameters in the model have to be in a certain range and satisfy
certain quantization conditions. In other words not every model can yield
geodesically complete cosmological solutions with zero size bounces. For
example, in the case of the flat FRW universe and in the absence of any
perturbations, the ratio of the parameters $b/c$ in the potential $V\left(
\sigma\right)  $ above must be in the range $-\frac{1}{4}\leq\left(
b/c\right)  \leq4$ and must be quantized as in Eq.(\ref{list}). This condition
on $b/c$ is relaxed in the presence of more parameters, such as curvature
$\left(  k=\pm1\right)  $ or radiation, but there is always one combination of
parameters and initial conditions that is quantized. Second, even with the
quantized parameters, initial conditions for $\phi\left(  \tau\right)
,s\left(  \tau\right)  $ must be synchronized with each other in order to
generate geodesically complete solutions in which $\left(  1-s^{2}/\phi
^{2}\right)  $ never changes sign. If initial conditions are not synchronized,
then $\left(  1-s^{2}/\phi^{2}\right)  $ will change sign and all solutions
will be geodesically incomplete; but this is what we want to avoid, and on
this basis we consider the solutions with synchronized boundary conditions,
namely only the geodesically complete ones, as being those that provide a
fuller story of cosmology.

We have also found exact analytic solutions, that obey
$\phi^{2}-s^{2}>0$ at all times, in which the initial conditions
\textit{need not be synchronized or the parameters of the model need
not be quantized}. Such solutions occur in the presence of spacial
curvature in the closed universe $k=1$, and provide a cyclic
cosmology where the universe bounces at a \textit{finite size. }For
this to be possible the curvature needs to be large enough to
compete with the potential energy $V\left(  \sigma\right) ,$ as will
be discussed in sections (\ref{bounceB},\ref{bounce2}).

Since $\left(  1-s^{2}/\phi^{2}\right)  $ never changes sign for such
solutions the physics at all times is compatible with Einstein's gravity,
since then one can indeed choose the Einstein gauge, Eq.(\ref{Egauge}), at all
times in such a universe.

If one takes the point of view that the theory is \textit{defined} directly in
the Einstein frame in terms of $a,\sigma$, as in Eq.(\ref{theory}), then the
$\phi,s$ configurations in which $\left(  1-s^{2}/\phi^{2}\right)  $ changes
sign is a spurious outcome of the parametrization in terms of $\phi,s$ in
Eq.(\ref{initialS}). In that case all field configurations in which $\left(
1-s^{2}/\phi^{2}\right)  $ is negative are excluded by definition. If one also
requires geodesic completeness then the solutions we present below are the
only ones that satisfy the criteria.

This avoids the question of what happens to the physics for those solutions
that \textit{are geodesically complete} in a more general sense than the
Einstein frame, by allowing $\left(  1-s^{2}/\phi^{2}\right)  $ to change
sign. If the theory is defined at a more fundamental level (as in 2T-physics,
or as in the colliding branes scenario) in which one would accept all the
consequences of the action in the $\phi,s$ version of Eq.(\ref{initialS}),
then one must investigate the properties of those solutions as well. What we
do know from our explicit solutions \cite{inflationBC}, in the cases in which
initial conditions are not synchronized or parameters are not quantized, is
that the quantity $\left(  \phi^{2}-s^{2}\right)  $ does not remain negative
after switching sign, but oscillates back to positive. So, it appears that the
universe recovers from antigravity and comes back to a period of time with
ordinary gravity. However, if allowed to continue its motion in complete
geodesics, the sign changes back and forth again and again. Perhaps the
physics appears to be all wrong during the time periods (or more generally
spacetime patches) where $\left(  \phi^{2}-s^{2}\right)  $ is negative, but we
don't really know the physical cosmological consequences of such solutions for
our own universe. We think that it would be interesting to find out eventually
the physical viability and meaning in cosmology of the more general
geodesically complete solutions allowed by the action Eq.(\ref{initialS}). So
we will not throw away yet the generic solutions which were included in the
list in (\cite{inflationBC}), nor will we settle the associated physics
questions in this paper. So, at a less ambitious level, for the moment we
concentrate only on the \textit{geodesically complete} cases that
\textit{also} satisfy $\phi^{2}-s^{2}\geq0,$ as required by the action
Eq.(\ref{theory}).

\section{Analytic solutions \label{gamma}}

To analyze the model in the $\phi,s$ version we find it useful to gauge fix
the Weyl symmetry in other forms. A very useful gauge is to choose the
conformal factor of the metric to be $1.$ We will call this the $\gamma
$-gauge. In this gauge we will denote the fields with a label $\gamma,$ such
as $\phi_{\gamma},s_{\gamma},g_{\gamma}^{\mu\nu}.$ Then, the Robertson-Walker
metric in this gauge loses the scale factor since $a_{\gamma}=1$
\begin{equation}
ds_{\gamma}^{2}=-d\tau^{2}+\frac{dr^{2}}{1-kr^{2}/r_{0}^{2}}+r^{2}\left(
d\theta^{2}+\sin^{2}\theta d\phi^{2}\right)  ,
\end{equation}
and its curvature is a constant given by
\begin{equation}
R\left(  g_{\gamma}\right)  =6K,~\text{with }K\equiv\frac{k}{r_{0}^{2}%
},\;k=0,\pm1.
\end{equation}
In this $\gamma$-gauge there is no scale factor for the universe, but now both
$\phi_{\gamma}\left(  x\right)  ,s_{\gamma}\left(  x\right)  $ are dynamical
variables, with $\phi_{\gamma}$ having the wrong sign in the kinetic term. The
advantage of this gauge is that the dynamics of $\phi_{\gamma},s_{\gamma}$
become much simpler and can be solved exactly in certain cases. After
obtaining the solution one can transform back to the Einstein gauge to find
$a\left(  \tau\right)  ,\sigma\left(  \tau\right)  .$ For the case of only
time dependent fields the gauge fixed form of the action (\ref{initialS}) is
\begin{equation}
L=\frac{1}{2}\left(  -\dot{\phi}_{\gamma}^{2}+\dot{s}_{\gamma}^{2}\right)
-\frac{K}{2}\left(  -\phi_{\gamma}^{2}+s_{\gamma}^{2}\right)  -\phi
^{4}f\left(  \frac{s}{\phi}\right)  . \label{Lfs}%
\end{equation}
But one should also remember that $\tau$ reparameterization symmetry of
general relativity requires the vanishing Hamiltonian constraint (this is the
$G_{00}=T_{00}$ Einstein equation)
\begin{equation}
H=\frac{1}{2}\left(  -p_{\phi}^{2}+p_{s}^{2}\right)  +\frac{K}{2}\left(
-\phi_{\gamma}^{2}+s_{\gamma}^{2}\right)  +\phi^{4}f\left(  \frac{s}{\phi
}\right)  =0, \label{Hconstraint}%
\end{equation}
where the canonical momenta are $p_{\phi}=-\dot{\phi}_{\gamma}$ and
$p_{s}=\dot{s}_{\gamma}.$ The negative norm ghost is eliminated because of
this constraint on the phase space $\left(  \phi_{\gamma},s_{\gamma},p_{\phi
},p_{s}\right)  $. The Wheeler-deWitt (WdW) equation of our theory
$H\Psi\left(  \phi,s\right)  =0$ takes an interesting form in the $\phi,s$
basis
\begin{equation}
\left(  \frac{1}{2}\left(  \partial_{\phi}^{2}-\partial_{s}^{2}\right)
+\frac{K}{2}\left(  -\phi_{\gamma}^{2}+s_{\gamma}^{2}\right)  +\phi
^{4}f\left(  \frac{s}{\phi}\right)  \right)  \Psi\left(  \phi,s\right)  =0.
\label{WdWeq}%
\end{equation}

As a side remark, note that for $K>0$ (closed universe) and in the absence of
the potential, $\phi^{4}f(\frac{s}{\phi})=0,$ the system in Eqs.(\ref{Lfs}%
-\ref{WdWeq}) describes the SO$\left(  1,1\right)  $ Lorentz symmetric
relativistic harmonic oscillator in 1+1 dimensions, with $\left(
\phi,s\right)  $ representing the (\textquotedblleft time\textquotedblright,
\textquotedblleft space\textquotedblright) directions respectively. As in
other cases of harmonic oscillator in several dimensions, this system has a
larger hidden symmetry, which is SU$\left(  1,1\right)  \supset$SO$\left(
1,1\right)  $ in this case. The quantum version of the relativistic harmonic
oscillator (i.e. the WdW equation for $f\left(  s/\phi\right)  =0$) was
studied and solved exactly in sections VI, VII and Appendix of a recent paper
\cite{relativisticHO} by using unitary representations of SU$\left(
1,1\right)  .$ This may be taken as a toy model to begin a study of the WdW
equation for our case\footnote{The WdW equation $H\Psi\left(  \phi,s\right)
=0$ is satisfied by an infinite set of solutions of the relativistic harmonic
oscillator \cite{relativisticHO}$.$ These are $\Psi\left(  \phi,s\right)
=\sum_{n=0}^{\infty}c_{n}\psi_{n}\left(  \phi\right)  \psi_{n}\left(
s\right)  ,$ where the $c_{n}$ are arbitrary and $\psi_{n}\left(  \phi\right)
,\psi_{n}\left(  s\right)  $ are the standard 1-dimensional harmonic
oscillator wavefunctions that satisfy the eigenvalue equations $\frac{1}%
{2}\left(  -\partial_{\phi}^{2}+K\phi^{2}\right)  \psi_{n}\left(  \phi\right)
=E_{n}\psi_{n}\left(  \phi\right)  $ and $\frac{1}{2}\left(  -\partial_{s}%
^{2}+Ks^{2}\right)  \psi_{n}\left(  s\right)  =E_{n}\psi_{n}\left(  s\right)
,$ where $E_{n}=\sqrt{K}\left(  n+\frac{1}{2}\right)  .$ For all these
solutions $\Psi\left(  \phi,s\right)  $ has an overall gaussian factor
$\exp[-\frac{\sqrt{K}}{2}\left(  \phi^{2}+s^{2}\right)  ]$ times a polynomial.
This shows that the probability distribution $\left\vert \Psi\left(
\phi,s\right)  \right\vert ^{2}$ for these generic solutions is \textit{not
purely \textquotedblleft timelike\textquotedblright} (not $\phi^{2}>s^{2}$),
but rather it covers both \textquotedblleft timelike\textquotedblright\ and
\textquotedblleft spacelike\textquotedblright\ regions in $\left(
\phi,s\right)  $ space. This is not surprising since for generic boundary
conditions the classical equations also do not obey $\phi^{2}\left(
\tau\right)  -s^{2}\left(  \tau\right)  \geq0$ at all times. Only special
boundary conditions with synchronized phases for $\phi\left(  \tau\right)
,s\left(  \tau\right)  $ at $\tau=0$ can yield classical solutions that have
this property, as we have explained in the text. Similarly, the relativistic
harmonic oscillator has just one quantum state whose probability distribution
is concentrated in the timelike region $\phi^{2}\geq s^{2}$; has a damping
factor of the form $\exp\left(  -\left(  \phi^{2}-s^{2}\right)  \right)  $ and
vanishes on the \textquotedblleft lightcone\textquotedblright\ $\phi^{2}%
=s^{2}.$ This was given in the appendix of \cite{relativisticHO} (interchange
spacelike with timelike in that appendix). This solution is the
\textquotedblleft timelike singlet of SU$\left(  1,1\right)  $%
\textquotedblright\ \cite{relativisticHO}. There is also a separate
\textquotedblleft spacelike singlet of SU$\left(  1,1\right)  $%
\textquotedblright, while the other generic solutions correspond to a
superposition of other non-singlet unitary representations of SU$(1,1)$
\cite{relativisticHO}. Referring to the comments of the last two paragraphs in
section (\ref{geodC}), if the fundamental action is in the Einstein frame (as
in Eq.(\ref{theory})), then only the timelike SU$(1,1)$ singlet is acceptable
as a solution of the WdW equation. If, on the other hand, the starting point
is more general (as in action (\ref{initialS})), then in order to favor only
the solutions that are consistent with $\phi^{2}-s^{2}>0,$ the model should
have an additional ingredient. This could be an appropriate potential energy
term, effects of radiation, curvature, etc., or some appropriate constraint
that is natural in the model. This would then characterize the
\textquotedblleft right\textquotedblright\ model. \label{relHO}}. Of course,
we are interested in the full WdW equation, including the potential energy
$\phi^{4}f\left(  \frac{s}{\phi}\right)  $, radiation, and anisotropy, as will
be discussed elsewhere \cite{bcst1}.

We now turn to the classical equations of motion satisfied by $\phi_{\gamma
},s_{\gamma},$ including the potential energy. Such classical solutions
provide a semi-classical approximation to the WdW equation. We are interested
in an exactly solvable case so that we can study the issues we raised with
certainty. One of those exactly solvable cases corresponds to a special form
of the potential, namely $\phi^{4}f\left(  s/\phi\right)  =b\phi^{4}+cs^{4},$
that in turn corresponds to the hyperbolic potential $V\left(  \sigma\right)
$ given in Eq.(\ref{potential}). The equations of motion for $\phi_{\gamma
}\left(  \tau\right)  $ and $s_{\gamma}\left(  \tau\right)  $ are directly
obtained from the Lagrangian or Hamiltonian given above. But it is also
instructive to derive the equations directly from the Friedmann equations in
Eqs.(\ref{00}-\ref{eom sigma}) by using the following connection between the
$\gamma$-gauge and the Einstein gauge (derived in \cite{inflationBC})%
\begin{equation}
a^{2}=\frac{\kappa^{2}}{6}\left(  \phi_{\gamma}^{2}-s_{\gamma}^{2}\right)
,\;\;\sigma=\frac{\sqrt{6}}{\kappa}\frac{1}{2}\ln\left(  \frac{\phi_{\gamma
}+s_{\gamma}}{\phi_{\gamma}-s_{\gamma}}\right)  , \label{Etransform}%
\end{equation}
which gives
\begin{equation}
V\left(  \sigma\right)  =\frac{6^{2}}{\kappa^{4}}\frac{b\phi_{\gamma}%
^{4}+cs_{\gamma}^{4}}{\left(  \phi_{\gamma}^{2}-s_{\gamma}^{2}\right)  ^{2}}.
\end{equation}
Inserting these in the Friedmann equations we obtain the equations for
$\phi,s$ as follows%
\begin{align}
0  &  =\ddot{\phi}_{\gamma}-4b\phi_{\gamma}^{3}+K\phi_{\gamma},\label{f}\\
0  &  =\ddot{s}_{\gamma}+4cs_{\gamma}^{3}+Ks_{\gamma},\label{s}\\
0  &  =\left(  \frac{1}{2}\dot{\phi}_{\gamma}^{2}-b\phi_{\gamma}^{4}+\frac
{1}{2}K\phi_{\gamma}^{2}\right)  -\left(  \frac{1}{2}\dot{s}_{\gamma}%
^{2}+cs_{\gamma}^{4}+\frac{1}{2}Ks_{\gamma}^{2}\right)  . \label{E}%
\end{align}
The important observation is that the second order equations for $\phi,s$
decuple from each other, so they are exactly solvable. The third equation
simply states that the energy of the $\phi$ solution must be matched to the
energy of the $s$ solution $E_{\phi}=E_{s}$. Once the solution is obtained it
is transformed back to the Einstein frame by using Eqs.(\ref{Etransform}),
thus providing the desired solutions for $a\left(  \tau\right)  ,\sigma\left(
\tau\right)  $ in the Einstein frame. The general generic solutions of these
equations, for all possible ranges of the parameters and boundary conditions
are listed in \cite{inflationBC} for the $K=0$ case. The solutions are
expressed in terms of Jacobi elliptic functions\footnote{The Jacobi elliptic
functions that we need for our solutions are denoted as $sn\left(  z|m\right)
,cn\left(  z|m\right)  ,dn\left(  z|m\right)  .$ These are periodic functions
that have properties similar to trigonometric functions. The following
formulas \cite{Abramowitz} are directly useful to verify our solutions
explicitly. The derivative of Jacobi elliptic functions are given in terms of
expressions somewhat similar to those for trigonometric functions.%
\begin{align}
\frac{d}{dz}sn\left(  z|m\right)   &  =cn\left(  z|m\right)  \times dn\left(
z|m\right)  ,\\
\frac{d}{dz}cn\left(  z|m\right)   &  =-sn\left(  z|m\right)  \times dn\left(
z|m\right)  ,\\
\frac{d}{dz}dn\left(  z|m\right)   &  =-m\times sn\left(  z|m\right)  \times
cn\left(  z|m\right)  .
\end{align}
They also satisfy quadratic relations, such as%
\begin{equation}
\left(  sn\left(  z|m\right)  \right)  ^{2}+\left(  cn\left(  z|m\right)
\right)  ^{2}=1;\;\ m\left(  sn\left(  z|m\right)  \right)  ^{2}+\left(
dn\left(  z|m\right)  \right)  ^{2}=1.
\end{equation}
\label{jacobiE}} as given in \cite{inflationBC} and below.

Now we focus on the subset of solutions that satisfy the criteria we laid out.
When the spacial curvature of the FRW universe is zero (i.e. $k=0$), we found
that geodesically complete solutions, with $\left(  \phi^{2}-s^{2}\right)
\geq0$ at all times, can occur only when the ratio $b/c$ takes on the
following quantized values in the range $-\frac{1}{4}\leq\frac{b}{c}\leq4,$
with $c$ positive, $c>0,$ and
\begin{equation}
b=\left\{
\begin{array}
[c]{c}%
\frac{4c}{n^{4}}\\
0\\
-\frac{c}{\left(  n+1\right)  ^{4}}%
\end{array}
\right.  ,~\text{with }n=1,2,3,\cdots. \label{list}%
\end{equation}
Note that each value of $n$ defines a given model. The explicit solutions are
given in Eqs.(\ref{fszerok},\ref{b-zerok},\ref{b0zerok}) for $b>0,~b<0$ and
$b=0$ respectively. When additional parameters, such as curvature, radiation,
etc. are included in the model, then this quantization condition on the model
is relaxed, but still some combination of parameters and integration constants
must be quantized as we will discuss. Furthermore, if the spacial curvature is
sufficiently large, so that the curvature term in the action can compete with
the potential, then we find, geodesically complete, finite, bouncing solutions
of a cyclic universe where the bounce occurs at a minimum finite size of the
universe. For such cases there is no quantization condition on the parameters
of the model, but instead, the initial conditions on the fields must be within
a certain range defined by those parameters.

Before we give the mathematical details, we first explain how the physics is
easily captured by interpreting these decupled equations in terms of an analog
mechanical problem of a particle moving in a potential. In the case of $\phi$
the Hamiltonian is $H\left(  \phi\right)  =\frac{1}{2}\dot{\phi}^{2}+V\left(
\phi\right)  ,$ with $V\left(  \phi\right)  =\frac{1}{2}K\phi^{2}-b\phi^{4},$
while in the case of $s$ the Hamiltonian is $H\left(  s\right)  =\frac{1}%
{2}\dot{s}^{2}+V\left(  s\right)  ,$ with $V\left(  s\right)  =\frac{1}%
{2}K\phi^{2}+cs^{4}.$ According to Eq.(\ref{E}) the only acceptable solutions
for $\phi,s$ are the ones that satisfy
\begin{equation}
H\left(  \phi\left(  \tau\right)  \right)  =H\left(  s\left(  \tau\right)
\right)  =E\label{HH}%
\end{equation}
The corresponding potentials are depicted in Figures-(\ref{zero-k}%
,\ref{positive-k},\ref{negative-k}) for the cases of $k=0,\pm1.$ We have
included the cases of positive $b$ (heavy solid curve $V\left(  \phi\right)
$) and negative $b$ (dashed curve $V\left(  \phi\right)  $). We have drawn the
pictures for only positive $c$ (solid thin curve $V\left(  s\right)  $) while
for negative $c$ the $V\left(  s\right)  $ curve is reflected from the
horizontal axis in each figure.

These figures, combined with the physical intuition of a particle in
potential, capture the physical aspects of our solutions. We approach the
mathematical analysis systematically for each figure and investigate the
various ranges of the parameters $b,c,K$ and the integration parameter $E.$ We
will start with the simplest case of zero curvature and analyze it thoroughly
in section (\ref{k0section}). We will then discuss the positive/negative
curvature cases separately in sections (\ref{k+section},\ref{k-section})
respectively.

\section{The flat ($k=0$) FRW universe \label{k0section}}

\begin{figure}
[h]
\begin{center}
\includegraphics[
height=2.1162in,
width=3.3797in
]%
{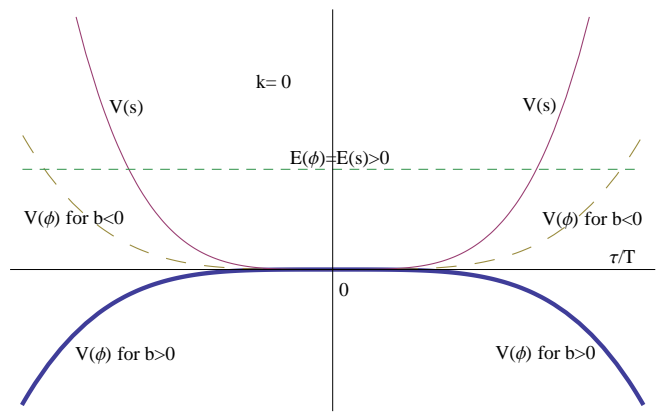}%
\caption{The flat FRW universe, $k=0.$}%
\label{zero-k}%
\end{center}
\end{figure}

We first discuss the flat case $\left(  k=0\right)  $ that has the fewest
parameters. For $k=0$ the only possible solutions are for $E_{s}=E_{\phi}>0$
as shown by the horizontal dashed line in Fig.(\ref{zero-k}). In the case of
$s,$ the potential is an infinite positive well ($V\left(  s\right)  =cs^{4}$
with $c>0$), therefore the particle is trapped in the well, and $s\left(
\tau\right)  $ oscillates back and forth between turning points $-s_{0}\left(
E_{s}\right)  <s\left(  \tau\right)  <s_{0}\left(  E_{s}\right)  $ given by
$cs_{0}^{4}=E_{s}$. In the case of $\phi,$ if $b$ is negative (dashed curve
$V\left(  \phi\right)  =-b\phi^{4}$ with $b<0$) its behavior is similar to the
one just described for $s,$ so $\phi\left(  \tau\right)  $ also oscillates
back and forth $-\phi_{0}\left(  E_{\phi}\right)  <\phi\left(  \tau\right)
<\phi_{0}\left(  E_{\phi}\right)  $ at the energy level $E_{\phi}=E_{s}$. But
if $b$ is positive, then the particle is in an inverted well (heavy curve
$V\left(  \phi\right)  =-b\phi^{4}$ with $b>0$). So, at the energy level
$E_{\phi}=E_{s},$ which is higher than the peak of the hill, the particle will
come up the hill from $\phi=-\infty,$ go over the top of the hill, and slide
down the hill to infinitely large positive values of $\phi.$ The trip may also
happen in reverse direction depending on initial conditions. It turns out that
the trip from $\phi=-\infty$ to $\phi=\infty$ is completed in a finite amount
of conformal time $\tau$, so allowing all possible values of proper time,
$\phi\left(  \tau\right)  $ repeats the trip periodically again and again by
jumping from $\phi=\infty$ to $\phi=-\infty.$ Such solutions (given
analytically in \cite{inflationBC}) solve all the equations but do not yet
address the sign of $\left(  \phi^{2}\left(  \tau\right)  -s^{2}\left(
\tau\right)  \right)  .$%

%TCIMACRO{\FRAME{fhFU}{3.3797in}{2.1162in}{0pt}{\Qcb{The flat FRW universe,
%$k=0.$}}{\Qlb{zero-k}}{cyclicpotentialvk0.eps}%
%width 3.3797in;  height 2.1162in;  depth 0pt;  original-width 3.333in;
%original-height 2.0781in;  cropleft "0";  croptop "1";  cropright "1";
%filename 'cyclicPotentialVk0.eps';file-properties "XNPEU";}}}%
%BeginExpansion
%EndExpansion

In order to have the oscillation amplitude of $\phi$ to be larger than the
amplitude of $s$ it is necessary to have $-\frac{c}{4}<b<4c$ (consistent with
the curves as drawn in the figure). The lower bound $-\frac{c}{4}<b$ for
negative $b,$ is partially understood from the figure which shows that the
turning point for $\phi$ should be at a greater distance from the origin as
compared to the turning point for $s.$ However the $\frac{1}{4}$ factor in
$-\frac{c}{4}<b$ and the upper bound $b<4c$ for positive $b,$ emerge from the
details of the solutions in Eqs.(\ref{fszerok},\ref{b-zerok},\ref{b0zerok}).
This is a constraint on the model. If the potential energy $V\left(
\sigma\right)  $ does not satisfy this property it will not be possible to
maintain $\phi^{2}\left(  \tau\right)  \geq s^{2}\left(  \tau\right)  $ at all
times. In addition, to insure $\phi^{2}\left(  \tau\right)  \geq s^{2}\left(
\tau\right)  ,$ we must $\left(  i\right)  $ synchronize the initial
conditions of the $\phi,s$ particles at the origin at $\tau=0,$ namely
$\phi\left(  0\right)  =s\left(  0\right)  =0,$ and $\left(  ii\right)  $ also
require that their periods are commensurate, so that at the time $\tau$ when
$\phi$ returns back to zero $s$ also returns to zero at the same time
(although $s$ could make several returns to zero in the meantime).
Commensurate periods can be arranged only by quantizing the parameter $b/c.$
This too is a condition on the model. If the quantization is not satisfied in
the model $\left(  \phi^{2}\left(  \tau\right)  -s^{2}\left(  \tau\right)
\right)  $ will change sign periodically as a function of time. But when the
potential $V\left(  \sigma\right)  $ satisfies the required conditions the
model yields geodesically complete solutions in which $\left(  \phi^{2}\left(
\tau\right)  -s^{2}\left(  \tau\right)  \right)  $ never changes sign, but
periodically touches zero, which corresponds to a big crunch smoothly followed
by a big bang. This is just the solution we sought as given in
Eqs.(\ref{fszerok},\ref{b-zerok},\ref{b0zerok}). We see that the model has to
be \textquotedblleft right\textquotedblright\ to be able to yield such a solution.

\subsection{$b>0$ case}

The solutions that satisfy this description are a subset of those in
\cite{inflationBC} and explicitly given by the following expressions. For
positive $b,c,$ the only geodesically complete solution occurs for the
quantized values of $b=4c/n^{4},$ with $n=1,2,3,\cdots,$ as follows
\begin{equation}
\phi_{\gamma}\left(  \tau\right)  =\frac{\kappa n}{\sqrt{48c}T}\frac{sn\left(
\frac{2\tau}{nT}|\frac{1}{2}\right)  }{1+cn\left(  \frac{2\tau}{nT}|\frac
{1}{2}\right)  },\;s_{\gamma}\left(  \tau\right)  =\frac{\kappa}{\sqrt{48c}%
T}\frac{sn\left(  \frac{\tau}{T}|\frac{1}{2}\right)  }{dn\left(  \frac{\tau
}{T}|\frac{1}{2}\right)  } \label{fszerok}%
\end{equation}
Here the Jacobi elliptic functions, $sn\left(  z|m\right)  $ etc. (see
footnote \ref{jacobiE}), appear only for the case of the parameter $m=1/2.$
The energy level $E_{\phi}=E_{s}$ is parametrized in terms of the parameter
$T$ which provides a scale for conformal time, as
\begin{equation}
E_{\phi}=E_{s}=\frac{1}{16cT^{4}},
\end{equation}
where $T$ (or $E_{s}=E_{\phi}$) is one of the integration parameters that
appears in integrating the differential equations. Note that this $T$ is
related to the overall factor in Eq.(\ref{fszerok}) that determines the
amplitude of oscillations of $s_{\gamma}\left(  \tau\right)  .$

\begin{figure}
[h]
\begin{center}
\includegraphics[
height=2.1015in,
width=3.3797in
]%
{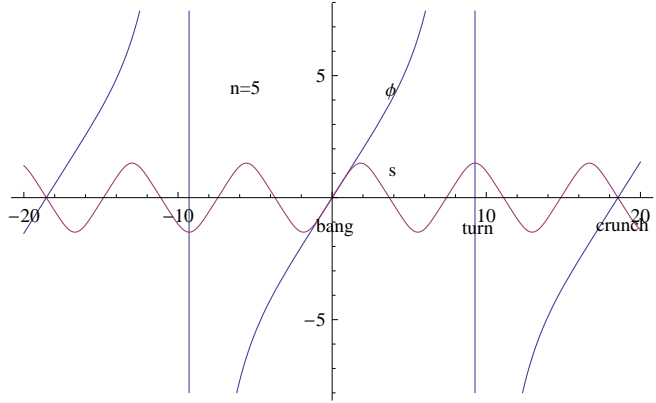}%
\caption{$\phi_{\gamma}\left(  \tau\right)  $ and $s_{\gamma}\left(
\tau\right)  $ plotted for $n=5,$ $\kappa=\sqrt{6},$ $T=1,$ $c=1/8.$}%
\label{fs5}%
\end{center}
\end{figure}

\begin{figure}
[h]
\begin{center}
\includegraphics[
height=2.0617in,
width=3.3797in
]%
{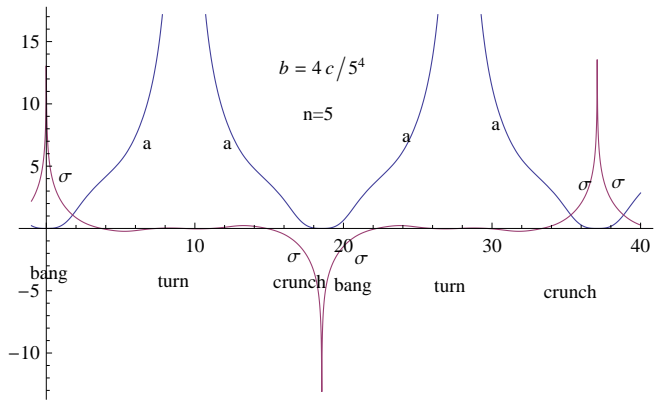}%
\caption{$a\left(  \tau\right)  $ and $\sigma\left(  \tau\right)  $ plotted
for $n=5,$ $\kappa=\sqrt{6},$ $T=1,$ $c=1/8.$}%
\label{as5}%
\end{center}
\end{figure}

It is easy to verify that these are solutions of Eqs.(\ref{f}-\ref{E}) by
using the properties of Jacobi elliptic functions given in footnote
\ref{jacobiE}. The plot of these functions in Fig.(\ref{fs5}) conveys their
periodic properties and show how $\phi_{\gamma}^{2}\left(  \tau\right)  \geq
s_{\gamma}^{2}\left(  \tau\right)  $ at all times.
%TCIMACRO{\FRAME{fhFU}{3.3797in}{2.1015in}{0pt}{\Qcb{$\phi_{\gamma}\left(
%\tau\right)  $ and $s_{\gamma}\left(  \tau\right)  $ plotted for $n=5,$
%$\kappa=\sqrt{6},$ $T=1,$ $c=1/8.$}}{\Qlb{fs5}}{cyclicfig.f_s_n5.eps}%
%width 3.3797in;  height 2.1015in;  depth 0pt;  original-width 3.333in;
%original-height 2.0626in;  cropleft "0";  croptop "1";  cropright "1";
%filename 'cyclicFig.f_s_n5.eps';file-properties "XNPEU";}}}%
%BeginExpansion
%EndExpansion
The quantum $n=5$ chosen for this figure corresponds to the ratio of the
periods of $\phi$ versus $s.$ The times at which $\phi$ and $s$ vanish
together $\phi_{\gamma}\left(  \tau\right)  =s_{\gamma}\left(  \tau\right)
=0$ are the only times when $\phi_{\gamma}^{2}\left(  \tau\right)  =s_{\gamma
}^{2}\left(  \tau\right)  ,$ at which point the universe goes through smoothly
from a big crunch to a big bang. At an intermediate time $\tau=\tau_{turn}$
the quantity $\phi_{\gamma}^{2}\left(  \tau\right)  -s_{\gamma}^{2}\left(
\tau\right)  $ attains a maximum; this is the turnaround point at which the
universe stops expanding and begins contracting. These features are seen in
Fig.(\ref{as5}) for the scale factor $a\left(  \tau\right)  ,$ and the scalar
$\sigma\left(  \tau\right)  $ which are given by Eqs.(\ref{Etransform}).%

%TCIMACRO{\FRAME{fhFU}{3.3797in}{2.0617in}{0pt}{\Qcb{$a\left(  \tau\right)  $
%and $\sigma\left(  \tau\right)  $ plotted for $n=5,$ $\kappa=\sqrt{6},$ $T=1,$
%$c=1/8.$}}{\Qlb{as5}}{cyclicfig_a+sigma_n5.eps}%
%width 3.3797in;  height 2.0617in;  depth 0pt;  original-width 3.333in;
%original-height 2.0228in;  cropleft "0";  croptop "1";  cropright "1";
%filename 'cyclicFig_a+sigma_n5.eps';file-properties "XNPEU";}}}%
%BeginExpansion
\begin{figure}
[hh]
\begin{center}
\includegraphics[
height=0.3096in,
width=5.2684in
]%
{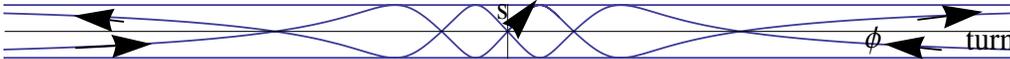}%
\caption{The arrow at the origin marks the crunch/bang moments and the arrows
at the ends mark the turnarounds.}%
\label{paramb+n5}%
\end{center}
\end{figure}

%EndExpansion
A parametric plot for $\phi_{\gamma}\left(  \tau\right)  ,s_{\gamma}\left(
\tau\right)  $ is given in Fig.(\ref{paramb+n5}), with $\phi$ on the
horizontal and $s$ on the vertical. This captures similar information to
Fig.(\ref{fs5}). It is for the model $b/c=4/n^{4}$ with $n=5$, which leads to
the 5 nodes in the figure. The time after the first node is a fast inflation
period, as seen also from Fig.(\ref{as5}). In a semi-classical approach to the
Wheeler-deWitt equation, the curve shows the region in $\left(  \phi,s\right)
$ space where the WdW wavefunction $\Psi\left(  \phi,s\right)  $ is expected
to have the largest probability. This is the unique curve (for the $n=5$
model) purely in the \textquotedblleft timelike\textquotedblright\ region
$\phi^{2}\left(  \tau\right)  -s^{2}\left(  \tau\right)  >0$ in $\left(
\phi,s\right)  $ space. The corresponding WdW wavefunction is the analog of
the \textquotedblleft timelike SU$\left(  1,1\right)  $
singlet\textquotedblright\ in footnote (\ref{relHO}).
%TCIMACRO{\FRAME{fhFU}{5.2684in}{0.3096in}{0pt}{\Qcb{The arrow at the origin
%marks the crunch/bang moments and the arrows at the ends mark the
%turnarounds.}}{\Qlb{paramb+n5}}{cyclicfig.parametric_f_s_n5.eps}%
%width 5.2684in;  height 0.3096in;  depth 0pt;  original-width 3.333in;
%original-height 0.1704in;  cropleft "0";  croptop "1";  cropright "1";
%filename 'cyclicFig.Parametric_f_s_n5.eps';file-properties "XNPEU";}%
%}}%
%BeginExpansion
%EndExpansion

\begin{figure}
[h]
\begin{center}
\includegraphics[
height=2.1015in,
width=3.3797in
]%
{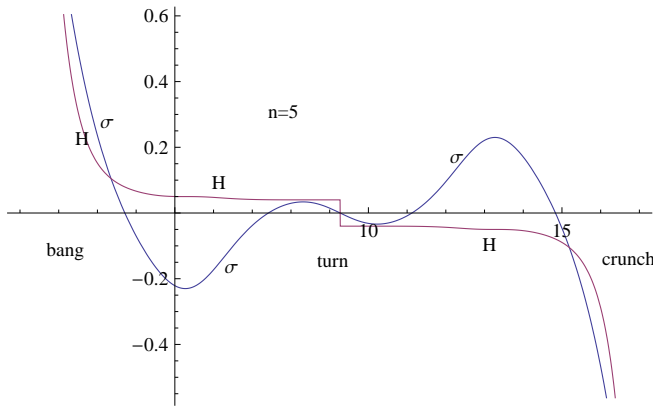}%
\caption{Hubble parameter $H\left(  \tau\right)  $ and $\sigma\left(
\tau\right)  $ for $n=5.$}%
\label{Hsig5}%
\end{center}
\end{figure}

\begin{figure}
[h]
\begin{center}
\includegraphics[
height=2.0617in,
width=3.3797in
]%
{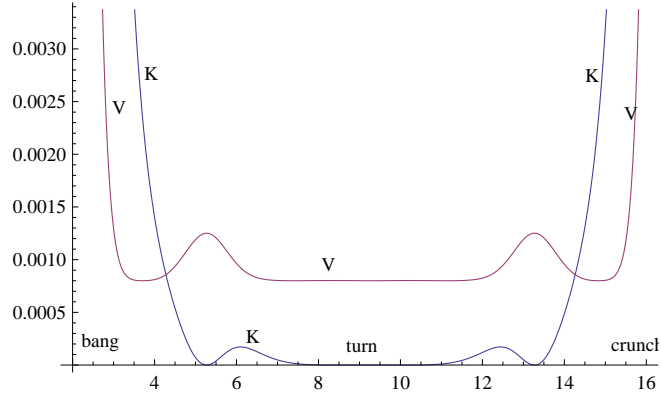}%
\caption{Kinetic $K\left(  \tau\right)  =\dot{\sigma}^{2}/2a^{2}$ and
potential energies $V\left(  \sigma\left(  \tau\right)  \right)  $ of the
$\sigma$ field, for $n=5$.}%
\label{KVa5}%
\end{center}
\end{figure}

\begin{figure}
[h]
\begin{center}
\includegraphics[
height=2.1162in,
width=3.3797in
]%
{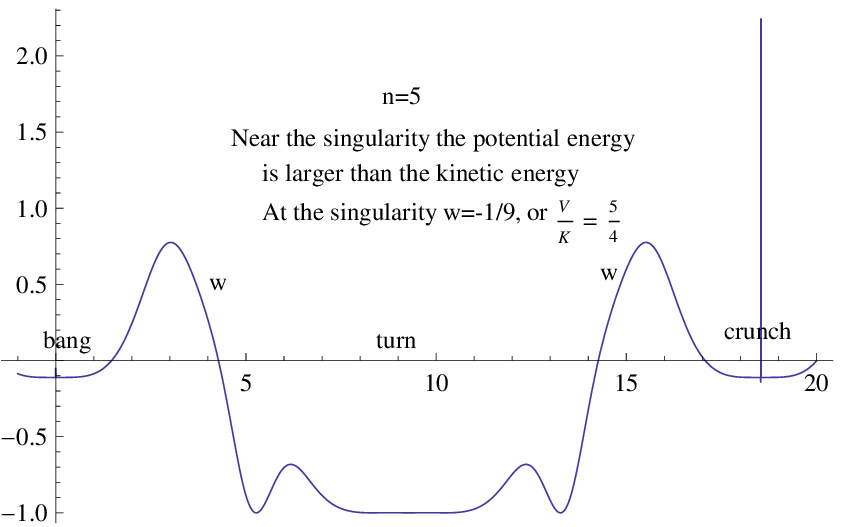}%
\caption{The equation of state $w=\left(  K-V\right)  /\left(  K+V\right)  ,$
for $n=5.$}%
\label{w5}%
\end{center}
\end{figure}

Other quantities of interest to convey the properties of the solution include
the Hubble parameter $H=\frac{\dot{a}}{a^{2}}$ (see footnote \ref{hubble}),
the kinetic energy of the $\sigma$ field, $K\left(  \tau\right)  =\frac
{\dot{\sigma}^{2}}{2a^{2}},$ its potential energy $V\left(  \sigma\left(
\tau\right)  \right)  $ and the equation of state parameter given by $w\left(
\tau\right)  =\left(  K\left(  \tau\right)  -V\left(  \tau\right)  \right)
/\left(  K\left(  \tau\right)  +V\left(  \tau\right)  \right)  $. Their plots
appear in Figs.(\ref{Hsig5},\ref{KVa5},\ref{w5}).
%TCIMACRO{\FRAME{fhFU}{3.3797in}{2.1015in}{0pt}{\Qcb{Hubble parameter $H\left(
%\tau\right)  $ and $\sigma\left(  \tau\right)  $ for $n=5.$}}{\Qlb{Hsig5}%
%}{cyclicfig.h_sigma_n5.eps}{\special{ language "Scientific Word";
%type "GRAPHIC";  maintain-aspect-ratio TRUE;  display "USEDEF";
%valid_file "F";  width 3.3797in;  height 2.1015in;  depth 0pt;
%original-width 3.333in;  original-height 2.0626in;  cropleft "0";
%filename 'cyclicFig.H_sigma_n5.eps';file-properties "XNPEU";}}}%
%BeginExpansion
%EndExpansion
%

%TCIMACRO{\FRAME{fhFU}{3.3797in}{2.0617in}{0pt}{\Qcb{Kinetic $K\left(
%\tau\right)  =\dot{\sigma}^{2}/2a^{2}$ and potential energies $V\left(
%\sigma\left(  \tau\right)  \right)  $ of the $\sigma$ field, for $n=5$.}%
%}{\Qlb{KVa5}}{cyclicfig.kva_n5.eps}{\special{ language "Scientific Word";
%type "GRAPHIC";  maintain-aspect-ratio TRUE;  display "USEDEF";
%valid_file "F";  width 3.3797in;  height 2.0617in;  depth 0pt;
%original-width 3.333in;  original-height 2.0228in;  cropleft "0";
%filename 'cyclicFig.KVa_n5.eps';file-properties "XNPEU";}}}%
%BeginExpansion
%EndExpansion
%

%TCIMACRO{\FRAME{fhFU}{3.3797in}{2.1162in}{0pt}{\Qcb{The equation of state
%$w=\left(  K-V\right)  /\left(  K+V\right)  ,$ for $n=5.$}}{\Qlb{w5}%
%}{cyclicfig.w_n5.eps}{\special{ language "Scientific Word";  type "GRAPHIC";
%width 3.3797in;  height 2.1162in;  depth 0pt;  original-width 3.333in;
%original-height 2.0781in;  cropleft "0";  croptop "1";  cropright "1";
%filename 'cyclicFig.w_n5.eps';file-properties "XNPEU";}}}%
%BeginExpansion
%EndExpansion

The Hubble parameter decreases from infinity at the big bang, quickly
approaching a constant at the turnaround (with a few small ripples depending
on $n$), switches to negative at turnaround and then slowly reaches negative
infinity at the big crunch.

The potential and kinetic energies of the $\sigma$ field are fairly close to
each other in magnitude most of the time. At the turnaround the kinetic energy
vanishes $K\left(  \tau_{turn}\right)  =0$ while the potential energy is a
constant $V\left(  \tau_{turn}\right)  =6^{2}b/\kappa^{4}$ (since $\sigma=0$
at turnaround). Both $K\left(  0\right)  $ and $V\left(  0\right)  $ are
infinite at the bang or crunch, but $V\left(  0\right)  $ is larger at the
singularity since $w\left(  0\right)  =-1/9$ as seen in Fig.(\ref{w5}).

The behavior of various quantities near the bang/crunch singularity is better
understood by studying the Taylor expansion near $\tau\rightarrow0$ for any
value of $n$ as follows%

\begin{align}
a\left(  \tau\right)   &  \rightarrow\frac{\kappa}{T\sqrt{48c}}\left(
\frac{\tau}{T}\right)  ^{3}\frac{\sqrt{4+n^{4}}}{2\sqrt{5}n^{2}}\left[
1+\frac{\left(  n^{2}-4\right)  }{60n^{2}}\left(  \frac{\tau}{T}\right)
^{4}+O\left(  \frac{\tau}{T}\right)  ^{8}\right] \\
\frac{\dot{a}\left(  \tau\right)  }{a\left(  \tau\right)  }  &  \rightarrow
\frac{1}{T}\left(  \frac{\tau}{T}\right)  ^{-1}\left[  3-\frac{n^{4}%
-4}{15n^{4}}\left(  \frac{\tau}{T}\right)  ^{4}+O\left(  \frac{\tau}%
{T}\right)  ^{8}\right] \\
H\left(  \tau\right)   &  \rightarrow\frac{\sqrt{48c}}{\kappa}\left(
\frac{\tau}{T}\right)  ^{-4}\left[  \frac{6\sqrt{5}n^{2}}{\sqrt{4+n^{4}}%
}-\frac{\sqrt{5}\left(  n^{4}-4\right)  }{30n^{2}\sqrt{4+n^{4}}}\left(
\frac{\tau}{T}\right)  ^{4}+O\left(  \frac{\tau}{T}\right)  ^{8}\right] \\
\sigma\left(  \tau\right)   &  \rightarrow\frac{\sqrt{6}}{\kappa}\left[
-\ln\left(  \frac{\tau}{T}\right)  ^{2}+\frac{1}{2}\ln\frac{80n^{4}}{4+n^{4}%
}+\frac{n^{4}-4}{240n^{4}}\left(  \frac{\tau}{T}\right)  ^{4}+O\left(
\frac{\tau}{T}\right)  ^{8}\right] \\
\dot{\sigma}\left(  \tau\right)   &  \rightarrow\frac{\sqrt{6}}{\kappa
T}\left(  \frac{\tau}{T}\right)  ^{-1}\left[  -2+\frac{n^{4}-4}{60n^{4}%
}\left(  \frac{\tau}{T}\right)  ^{4}+O\left(  \frac{\tau}{T}\right)
^{8}\right] \\
V\left(  \sigma\left(  \tau\right)  \right)   &  \rightarrow\frac{288c}%
{\kappa^{4}}\left(  \frac{\tau}{T}\right)  ^{-8}\left[  \frac{50n^{4}}{\left(
4+n^{4}\right)  }-\frac{5\left(  n^{4}-4\right)  }{3\left(  n^{4}+4\right)
}\left(  \frac{\tau}{T}\right)  ^{4}+O\left(  \frac{\tau}{T}\right)
^{8}\right] \\
K\left(  \sigma\left(  \tau\right)  \right)   &  \rightarrow\frac{288c}%
{\kappa^{4}}\left(  \frac{\tau}{T}\right)  ^{-8}\left[  \frac{40n^{4}}{\left(
4+n^{4}\right)  }+\frac{2\left(  n^{4}-4\right)  }{3\left(  n^{4}+4\right)
}\left(  \frac{\tau}{T}\right)  ^{4}+O\left(  \frac{\tau}{T}\right)
^{8}\right] \\
w\left(  \tau\right)   &  \rightarrow-\frac{1}{9}+\frac{2\left(
n^{4}-4\right)  }{81n^{4}}\left(  \frac{\tau}{T}\right)  ^{4}+O\left(
\frac{\tau}{T}\right)  ^{8}\nonumber
\end{align}
The last expression shows that $w=-1/9$ at the singularity for all values of
$n.$ This behavior seems to be surprising according to common lore.

We emphasize that this behavior near the singularity is only for our
geodesically complete analytic solutions that satisfy both the relative
quantization of their periods as well as the synchronization of the initial
conditions. If either of these is not satisfied (i.e. for non-geodesically
complete solutions in only the Einstein frame) the behavior near the
singularity is radically different. This behavior seems to be of measure zero
in the space of all solutions. In our next paper \cite{bcst1} we will further
analyze this issue including the effects of anisotropy and the quantum effects
via the Wheeler-deWitt equation.

\subsection{$b<0$ case}

We repeat the same type of analysis for $b<0$ which refers to
Fig.(\ref{zero-k}) with the dashed curve representing $V\left(  \phi\right)
.$ Geodesically complete solutions occur only if $b/c$ has one of the
quantized values $b/c=-1/\left(  n+1\right)  ^{4}$, with $n=1,2,3,\cdots.$
Then the unique solution is
\begin{equation}
\phi_{\gamma}\left(  \tau\right)  =\frac{\kappa\left(  n+1\right)  }%
{\sqrt{48c}T}\frac{sn\left(  \frac{\tau}{\left(  n+1\right)  T}|\frac{1}%
{2}\right)  }{dn\left(  \frac{\tau}{\left(  n+1\right)  T}|\frac{1}{2}\right)
},\;s_{\gamma}\left(  \tau\right)  =\frac{\kappa}{\sqrt{48c}T}\frac{sn\left(
\frac{\tau}{T}|\frac{1}{2}\right)  }{dn\left(  \frac{\tau}{T}|\frac{1}%
{2}\right)  }. \label{b-zerok}%
\end{equation}
We provide a few plots similar to the ones in the previous subsection for the
model $n=5.$ Despite some similarities, there are notable differences in the
behavior as compared to the $b>0$ case of the previous section as indicated in
the following comments.%

\[%
\begin{array}
[c]{cc}%
%TCIMACRO{\FRAME{itbpFU}{3.2404in}{2.015in}{0in}{\Qcb{FIG. 8: $\phi\left(
%\tau\right)  ,s\left(  \tau\right)  ,$ have finite amplitudes.}}{\Qlb{fs5b-}%
%}{cyclicfig.f_{s}_{n}5b-.eps}{\special{ language "Scientific Word";
%type "GRAPHIC";  maintain-aspect-ratio TRUE;  display "USEDEF";
%valid_file "F";  width 3.2404in;  height 2.015in;  depth 0in;
%original-width 3.333in;  original-height 2.0626in;  cropleft "0";
%filename 'cyclicFig.f_s_n5b-.eps';file-properties "XNPEU";}}}%
%BeginExpansion
{\parbox[b]{3.2404in}{\begin{center}
\includegraphics[
height=2.015in,
width=3.2404in
]%
{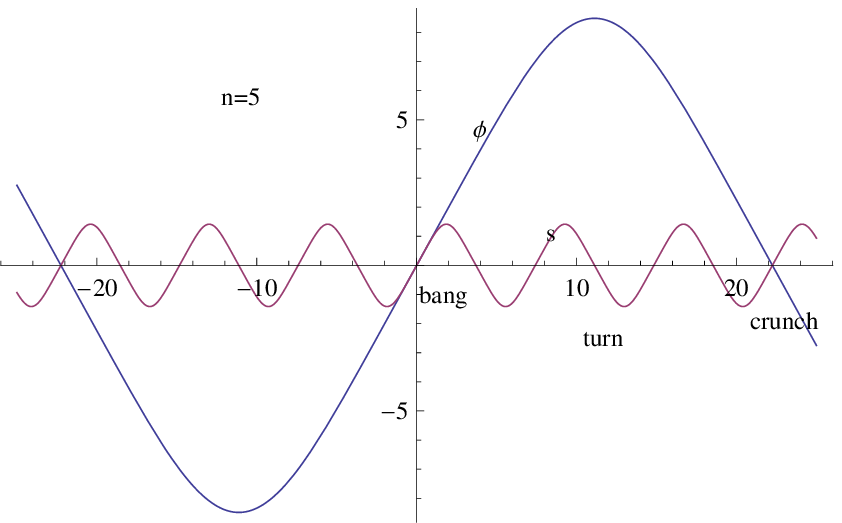}%
\\
FIG. 8: $\phi\left(  \tau\right)  ,s\left(  \tau\right)  ,$ have finite
amplitudes.
\end{center}}}%
%EndExpansion
&
%TCIMACRO{\FRAME{itbpFU}{3.2612in}{2.028in}{0in}{\Qcb{FIG. 9: $a\left(
%\tau\right)  $ has a finite maximum.}}{\Qlb{as5b-}}{cyclicfig_{a}%
%+sigma_{n}5b-.eps}{\special{ language "Scientific Word";  type "GRAPHIC";
%width 3.2612in;  height 2.028in;  depth 0in;  original-width 3.333in;
%original-height 2.0626in;  cropleft "0";  croptop "1";  cropright "1";
%filename 'cyclicFig_a+sigma_n5b-.eps';file-properties "XNPEU";}}}%
%BeginExpansion
{\parbox[b]{3.2612in}{\begin{center}
\includegraphics[
height=2.028in,
width=3.2612in
]%
{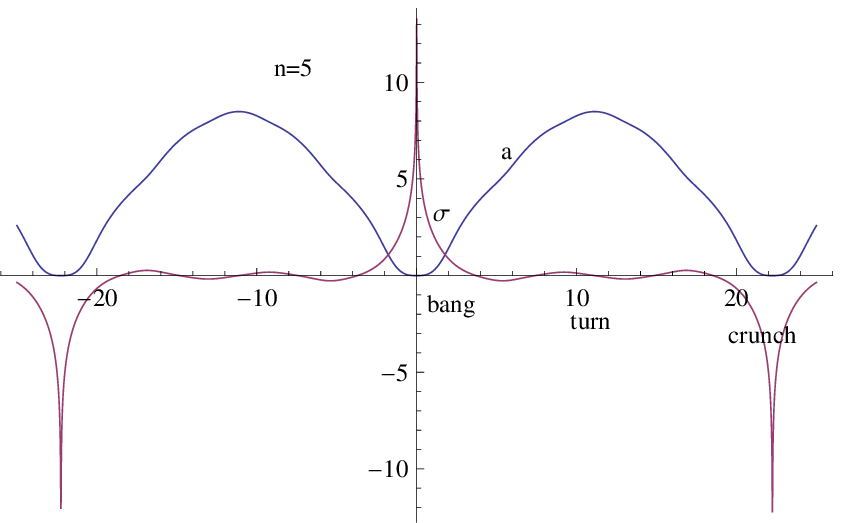}%
\\
FIG. 9: $a\left(  \tau\right)  $ has a finite maximum.
\end{center}}}%
%EndExpansion
\end{array}
\]
From Figures 8,9 we see that the universe grows up to a maximum finite size
before it turns around. Another way of plotting the information in Fig.8 is
the parametric plot for $\phi\left(  \tau\right)  ,s\left(  \tau\right)  $ in
Fig.(\ref{parametric_fs5b-}); note the 5 nodes corresponding to $n=5.$
\setcounter {figure} {9}
%TCIMACRO{\FRAME{fhFU}{3.3797in}{0.5898in}{0pt}{\Qcb{Crunch/bang is at the
%origin, turnaround at the edges.}}{\Qlb{parametric_fs5b-}}%
%{cyclicfig.parametric_f_s_n5b-.eps}{\special{ language "Scientific Word";
%type "GRAPHIC";  maintain-aspect-ratio TRUE;  display "USEDEF";
%valid_file "F";  width 3.3797in;  height 0.5898in;  depth 0pt;
%original-width 3.333in;  original-height 0.5587in;  cropleft "0";
%filename 'cyclicFig.Parametric_f_s_n5b-.eps';file-properties "XNPEU";}%
%}}%
%BeginExpansion
\begin{figure}
[h]
\begin{center}
\includegraphics[
height=0.5898in,
width=3.3797in
]%
{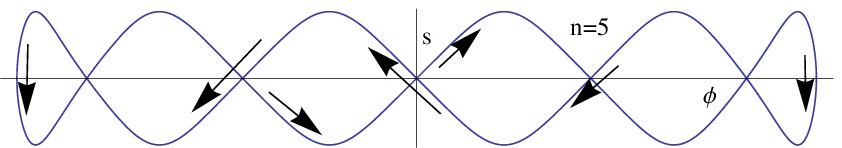}%
\caption{Crunch/bang is at the origin, turnaround at the edges.}%
\label{parametric_fs5b-}%
\end{center}
\end{figure}
%EndExpansion
As in the previous case, this figure is associated with the semiclassical
probability distribution of the Wheeler-deWitt wavefunction in $\left(
\phi,s\right)  $ space.%

\[%
\begin{array}
[c]{cc}%
%TCIMACRO{\FRAME{itbpFU}{3.2335in}{2.0116in}{0in}{\Qcb{FIG. 11: Behavior of
%$\sigma\left(  \tau\right)  ,H\left(  \tau\right)  ~$}}{\Qlb{Hsig5b-}%
%}{cyclicfig.h_{s}igma_{n}5b-.eps}{\special{ language "Scientific Word";
%type "GRAPHIC";  maintain-aspect-ratio TRUE;  display "USEDEF";
%valid_file "F";  width 3.2335in;  height 2.0116in;  depth 0in;
%original-width 3.333in;  original-height 2.0626in;  cropleft "0";
%filename 'cyclicFig.H_sigma_n5b-.eps';file-properties "XNPEU";}}}%
%BeginExpansion
{\parbox[b]{3.2335in}{\begin{center}
\includegraphics[
height=2.0116in,
width=3.2335in
]%
{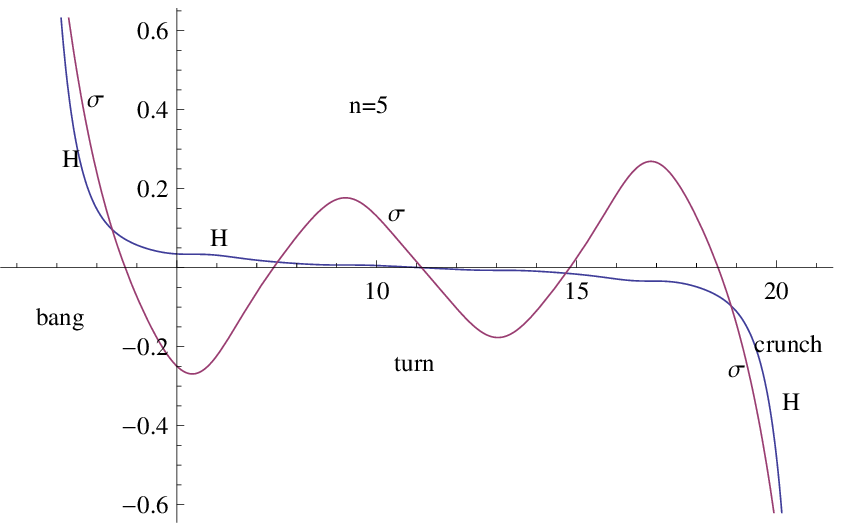}%
\\
FIG. 11: Behavior of $\sigma\left(  \tau\right)  ,H\left(  \tau\right)  ~$
\end{center}}}%
%EndExpansion
&
%TCIMACRO{\FRAME{itbpFU}{3.2612in}{2.028in}{0in}{\Qcb{FIG. 12: Temporary
%inflation periods.}}{\Qlb{H_{a}cc5b-}}{cyclicfig.h_{a}cc_{n}5b-.eps}%
%width 3.2612in;  height 2.028in;  depth 0in;  original-width 3.333in;
%original-height 2.0626in;  cropleft "0";  croptop "1";  cropright "1";
%filename 'cyclicFig.H_acc_n5b-.eps';file-properties "XNPEU";}}}%
%BeginExpansion
{\parbox[b]{3.2612in}{\begin{center}
\includegraphics[
height=2.028in,
width=3.2612in
]%
{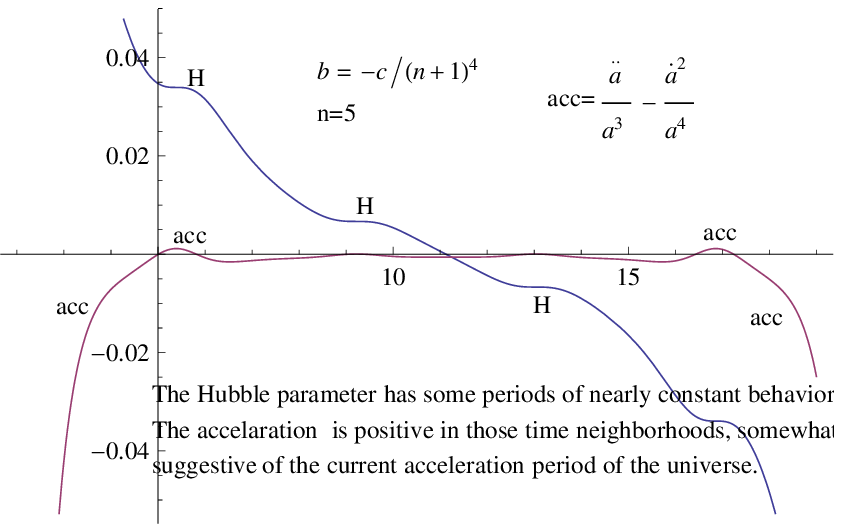}%
\\
FIG. 12: Temporary inflation periods.
\end{center}}}%
%EndExpansion
\end{array}
\]
From Figs.(11,12) we see that there are temporary inflation periods during
which $H\left(  \tau\right)  $ is temporarily almost a constant, and the
acceleration\footnote{The acceleration is defined in the Einstein frame as
$\frac{d^{2}a_{E}\left(  t\right)  }{dt^{2}}.$ This may be written in terms of
conformal time as $\frac{1}{a}\partial_{\tau}\left(  \frac{1}{a}\partial
_{\tau}a\right)  .$ For the purpose of the plot we have defined the quantity
\textquotedblleft acc\textquotedblright, as the acceleration divided by an
extra factor of $a$.} is positive, as seen in Fig.12. The number of such
temporary acceleration periods is determined by $n.$

It is interesting to speculate on whether this could be a mechanism to explain
the current accelerated inflation state of our universe; namely could it be
that we currently are in such a period which is inflationary only temporarily
on the scale of the lifetime of the universe?%

\[%
\begin{array}
[c]{cc}%
%TCIMACRO{\FRAME{itbpFU}{3.2059in}{2.0081in}{0in}{\Qcb{FIG.13: Energy
%components of $\sigma$ field.}}{\Qlb{KV5b-}}{cyclicfig.kva_{n}5b-.eps}%
%width 3.2059in;  height 2.0081in;  depth 0in;  original-width 3.333in;
%original-height 2.0781in;  cropleft "0";  croptop "1";  cropright "1";
%filename 'cyclicFig.KVa_n5b-.eps';file-properties "XNPEU";}}}%
%BeginExpansion
{\parbox[b]{3.2059in}{\begin{center}
\includegraphics[
height=2.0081in,
width=3.2059in
]%
{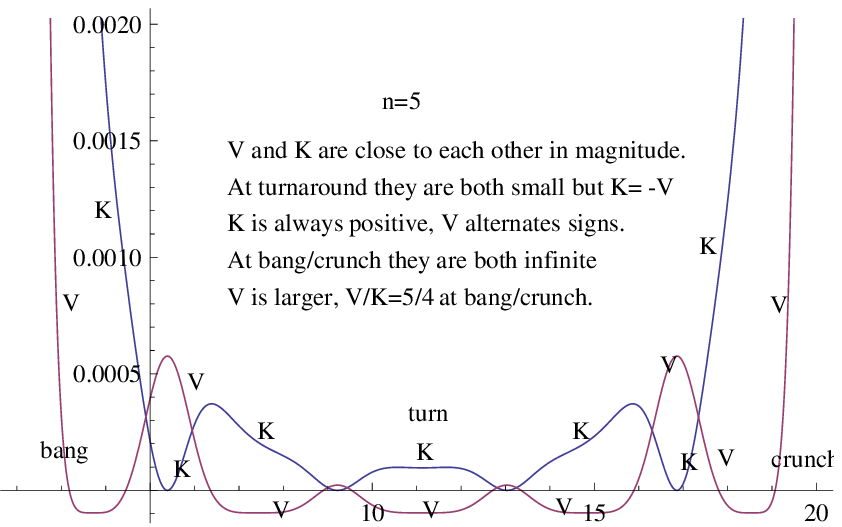}%
\\
FIG.13: Energy components of $\sigma$ field.
\end{center}}}%
%EndExpansion
&
%TCIMACRO{\FRAME{itbpFU}{3.2543in}{2.0237in}{0in}{\Qcb{FIG.14: Equation of
%state $w\left(  \tau\right)  .$}}{\Qlb{w5b-}}{cyclicfig.w_{n}5b-.eps}%
%width 3.2543in;  height 2.0237in;  depth 0in;  original-width 3.333in;
%original-height 2.0626in;  cropleft "0";  croptop "1";  cropright "1";
%filename 'cyclicFig.w_n5b-.eps';file-properties "XNPEU";}}}%
%BeginExpansion
{\parbox[b]{3.2543in}{\begin{center}
\includegraphics[
height=2.0237in,
width=3.2543in
]%
{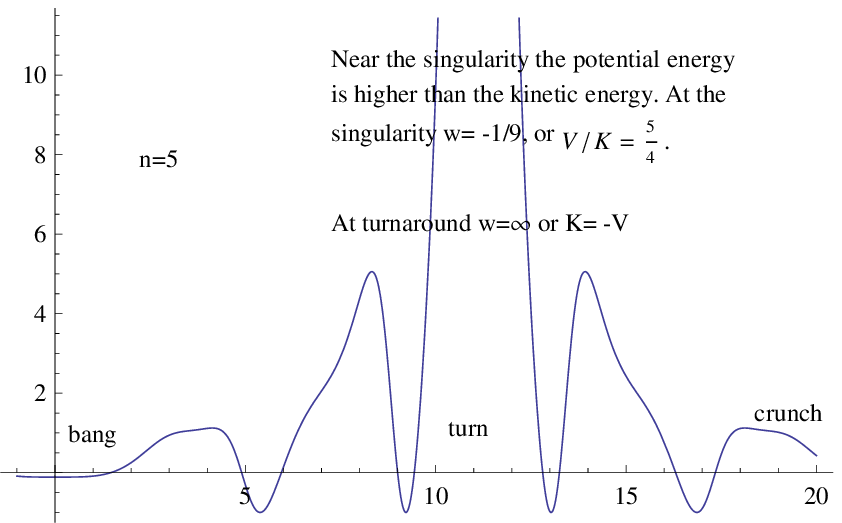}%
\\
FIG.14: Equation of state $w\left(  \tau\right)  .$
\end{center}}}%
%EndExpansion
\end{array}
\]
The energy of the $\sigma$ field is small except near the bang/crunch where it
is infinite. The equation of state $w\left(  \tau\right)  $ grows to infinity
at turnaround $w\left(  \tau_{turn}\right)  =\infty$, while it takes the value
$w\left(  0\right)  =-1/9$ at the bang/crunch where $V\left(  0\right)
/K\left(  0\right)  =5/4$, while $V\left(  0\right)  ,K\left(  0\right)  $ are
both infinite.

The behavior of various quantities near the bang/crunch singularity is given
by the Taylor expansion near $\tau\rightarrow0$ for any value of $n$ as follows%

\begin{align}
a\left(  \tau\right)   &  \rightarrow\frac{\kappa}{T\sqrt{48c}}\left(
\frac{\tau}{T}\right)  ^{3}\frac{\sqrt{\left(  n+1\right)  ^{4}-1}}{2\sqrt
{5}\left(  n+1\right)  ^{2}}\left[  1-\left(  \frac{\tau}{T}\right)  ^{4}%
\frac{\left(  n+1\right)  ^{4}+1}{60\left(  n+1\right)  ^{4}}+O\left(
\frac{\tau}{T}\right)  ^{8}\right] \\
\frac{\dot{a}\left(  \tau\right)  }{a\left(  \tau\right)  }  &  \rightarrow
\frac{1}{T}\left(  \frac{\tau}{T}\right)  ^{-1}\left[  3-\frac{\left(
n+1\right)  ^{4}+1}{15\left(  n+1\right)  ^{4}}\left(  \frac{\tau}{T}\right)
^{4}+O\left(  \frac{\tau}{T}\right)  ^{8}\right] \\
H\left(  \tau\right)   &  \rightarrow\frac{\sqrt{48c}}{\kappa}\left(
\frac{\tau}{T}\right)  ^{-4}\frac{6\sqrt{5}\left(  n+1\right)  ^{2}}%
{\sqrt{\left(  n+1\right)  ^{4}-1}}\left[  1-\frac{\left(  n+1\right)  ^{4}%
+1}{180\left(  n+1\right)  ^{4}}\left(  \frac{\tau}{T}\right)  ^{4}+O\left(
\frac{\tau}{T}\right)  ^{8}\right] \\
\sigma\left(  \tau\right)   &  \rightarrow\frac{\sqrt{6}}{\kappa}\left[
-\ln\left(  \frac{\tau}{T}\right)  ^{2}+\frac{1}{2}\ln\frac{80\left(
n+1\right)  ^{4}}{\left(  n+1\right)  ^{4}-1}+\frac{\left(  n+1\right)
^{4}+1}{240\left(  n+1\right)  ^{4}}\left(  \frac{\tau}{T}\right)
^{4}+O\left(  \frac{\tau}{T}\right)  ^{8}\right] \\
\dot{\sigma}\left(  \tau\right)   &  \rightarrow\frac{\sqrt{6}}{\kappa
T}\left(  \frac{\tau}{T}\right)  ^{-1}\left[  -2+\frac{\left(  n+1\right)
^{4}+1}{60\left(  n+1\right)  ^{4}}\left(  \frac{\tau}{T}\right)
^{4}+O\left(  \frac{\tau}{T}\right)  ^{8}\right] \\
V\left(  \sigma\left(  \tau\right)  \right)   &  \rightarrow\frac{288c}%
{\kappa^{4}}\left(  \frac{\tau}{T}\right)  ^{-8}\left[  \frac{50\left(
n+1\right)  ^{4}}{\left(  \left(  n+1\right)  ^{4}-1\right)  }-\frac{5}%
{3}\frac{\left(  n+1\right)  ^{4}+1}{\left(  \left(  n+1\right)
^{4}-1\right)  }\left(  \frac{\tau}{T}\right)  ^{4}+O\left(  \frac{\tau}%
{T}\right)  ^{8}\right] \\
K\left(  \sigma\left(  \tau\right)  \right)   &  \rightarrow\frac{288c}%
{\kappa^{4}}\left(  \frac{\tau}{T}\right)  ^{-8}\left[  \frac{40\left(
n+1\right)  ^{4}}{\left(  \left(  n+1\right)  ^{4}-1\right)  }+\frac{2}%
{3}\frac{\left(  n+1\right)  ^{4}+1}{\left(  \left(  n+1\right)
^{4}-1\right)  }\left(  \frac{\tau}{T}\right)  ^{4}+O\left(  \frac{\tau}%
{T}\right)  ^{8}\right] \\
w\left(  \tau\right)   &  \rightarrow-\frac{1}{9}+\frac{2}{81}\frac{\left(
n+1\right)  ^{4}+1}{60\left(  n+1\right)  ^{4}}\left(  \frac{\tau}{T}\right)
^{4}+O\left(  \frac{\tau}{T}\right)  ^{8}\nonumber
\end{align}

\subsection{$b=0$ case}

Finally, for vanishing $b=0,$ the solution corresponds to the $n\rightarrow
\infty$ limit of either the positive or negative $b$ branches, and is given
by
\begin{equation}
\phi_{\gamma}\left(  \tau\right)  =\frac{\kappa}{\sqrt{48c}T}\frac{\tau}%
{T},\;\;s_{\gamma}\left(  \tau\right)  =\frac{\kappa}{\sqrt{48c}T}%
\frac{sn\left(  \frac{\tau}{T}|\frac{1}{2}\right)  }{dn\left(  \frac{\tau}%
{T}|\frac{1}{2}\right)  }. \label{b0zerok}%
\end{equation}
We provide a few plots similar to the ones in the previous subsections.%

\begin{equation}%
\begin{array}
[c]{cc}%
%TCIMACRO{\FRAME{itbpFU}{3.0113in}{1.8853in}{0in}{\Qcb{FIG.15: $\phi\left(
%\tau\right)  $ grows linearly with $\tau.$}}{}{cyclicfig.f_{s}_{b}%
%0.eps}{\special{ language "Scientific Word";  type "GRAPHIC";
%width 3.0113in;  height 1.8853in;  depth 0in;  original-width 3.333in;
%original-height 2.0781in;  cropleft "0";  croptop "1";  cropright "1";
%filename 'cyclicFig.f_s_b0.eps';file-properties "XNPEU";}}}%
%BeginExpansion
{\parbox[b]{3.0113in}{\begin{center}
\includegraphics[
height=1.8853in,
width=3.0113in
]%
{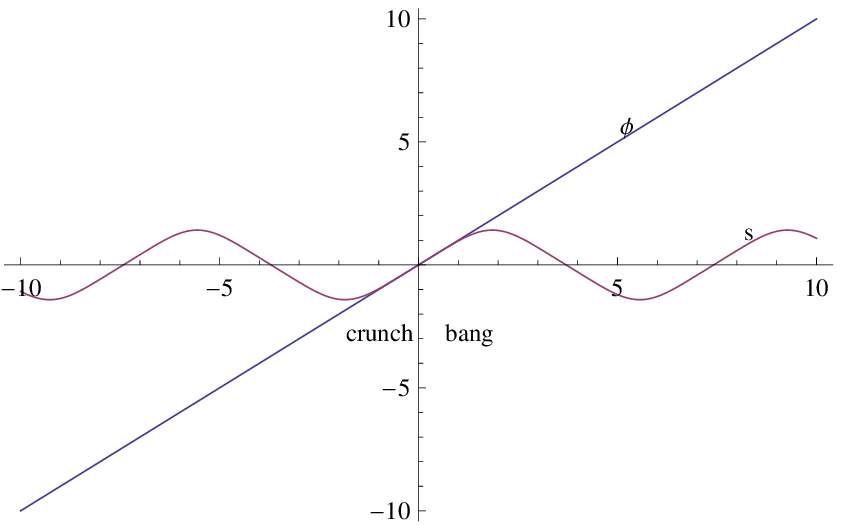}%
\\
FIG.15: $\phi\left(  \tau\right)  $ grows linearly with $\tau.$
\end{center}}}%
%EndExpansion
&
%TCIMACRO{\FRAME{itbpFU}{3.0251in}{1.9329in}{0in}{\Qcb{FIG.16: There is a
%single crunch/bang.}}{}{cyclicfig_{a}+sigma_{n}5b0.eps}%
%width 3.0251in;  height 1.9329in;  depth 0in;  original-width 3.333in;
%original-height 2.1205in;  cropleft "0";  croptop "1";  cropright "1";
%filename 'cyclicFig_a+sigma_n5b0.eps';file-properties "XNPEU";}}}%
%BeginExpansion
{\parbox[b]{3.0251in}{\begin{center}
\includegraphics[
height=1.9329in,
width=3.0251in
]%
{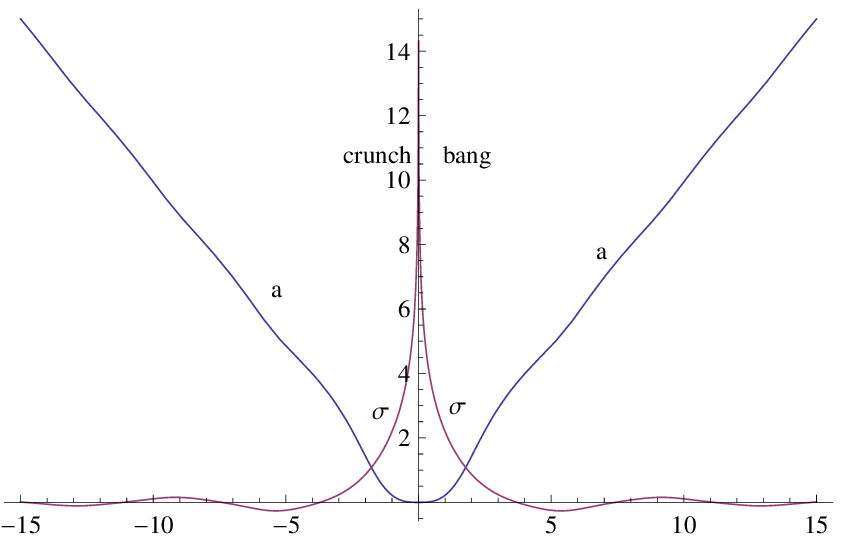}%
\\
FIG.16: There is a single crunch/bang.
\end{center}}}%
%EndExpansion
\end{array}
\end{equation}

\setcounter {figure} {16}%
%TCIMACRO{\FRAME{fhFU}{3.3797in}{0.5068in}{0pt}{\Qcb{Parametric plot equivalent
%to Fig.15.}}{}{cyclicfig.parametric_f_s_n5b0.eps}%
%width 3.3797in;  height 0.5068in;  depth 0pt;  original-width 3.333in;
%original-height 0.4765in;  cropleft "0";  croptop "1";  cropright "1";
%filename 'cyclicFig.Parametric_f_s_n5b0.eps';file-properties "XNPEU";}%
%}}%
%BeginExpansion
\begin{figure}
[h]
\begin{center}
\includegraphics[
height=0.5068in,
width=3.3797in
]%
{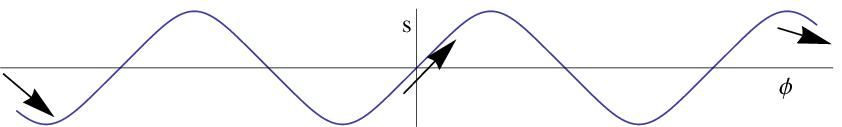}%
\caption{Parametric plot equivalent to Fig.15.}%
\end{center}
\end{figure}
%EndExpansion
These plots correspond to the $n=\infty$ limit of the previous plots for
either $b>0$ or $b<0.$ Therefore their interpretation is similar to the
discussion above%

\begin{equation}%
\begin{array}
[c]{cc}%
%TCIMACRO{\FRAME{itbpFU}{3.0251in}{1.9061in}{0in}{\Qcb{FIG.18: $H\left(
%\tau\right)  ,$ $\sigma\left(  \tau\right)  $ decrease.}}{}{cyclicfig.h_{s}%
%igma_{n}5b0.eps}{\special{ language "Scientific Word";  type "GRAPHIC";
%width 3.0251in;  height 1.9061in;  depth 0in;  original-width 3.333in;
%original-height 2.0903in;  cropleft "0";  croptop "1";  cropright "1";
%filename 'cyclicFig.H_sigma_n5b0.eps';file-properties "XNPEU";}}}%
%BeginExpansion
{\parbox[b]{3.0251in}{\begin{center}
\includegraphics[
height=1.9061in,
width=3.0251in
]%
{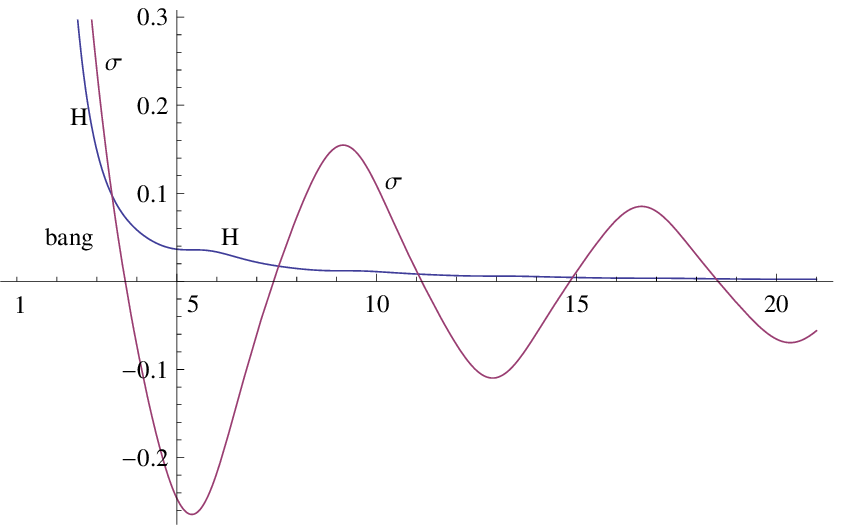}%
\\
FIG.18: $H\left(  \tau\right)  ,$ $\sigma\left(  \tau\right)  $ decrease.
\end{center}}}%
%EndExpansion
&
%TCIMACRO{\FRAME{itbpFU}{3.0597in}{1.9164in}{0in}{\Qcb{FIG.19: Temporary
%acceleration periods.}}{}{cyclicfig.h_{a}cc_{n}5b0.eps}%
%width 3.0597in;  height 1.9164in;  depth 0in;  original-width 3.333in;
%original-height 2.0781in;  cropleft "0";  croptop "1";  cropright "1";
%filename 'cyclicFig.H_acc_n5b0.eps';file-properties "XNPEU";}}}%
%BeginExpansion
{\parbox[b]{3.0597in}{\begin{center}
\includegraphics[
height=1.9164in,
width=3.0597in
]%
{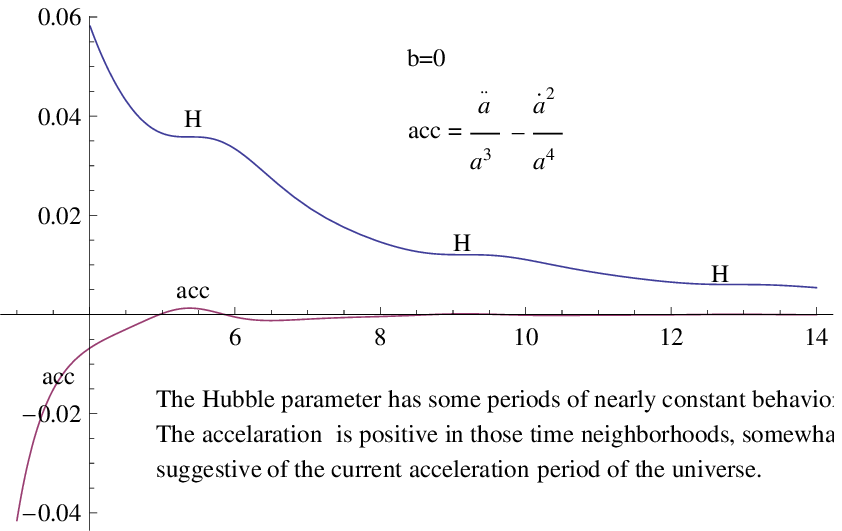}%
\\
FIG.19: Temporary acceleration periods.
\end{center}}}%
%EndExpansion
\end{array}
\end{equation}

The temporary acceleration periods persist.%

\begin{equation}%
\begin{array}
[c]{cc}%
%TCIMACRO{\FRAME{itbpFU}{2.9905in}{1.849in}{0in}{\Qcb{FiG.20: The energy of
%$\sigma$ field decreases.}}{}{cyclicfig.kva_{n}5b0.eps}%
%width 2.9905in;  height 1.849in;  depth 0in;  original-width 3.333in;
%original-height 2.0505in;  cropleft "0";  croptop "1";  cropright "1";
%filename 'cyclicFig.KVa_n5b0.eps';file-properties "XNPEU";}}}%
%BeginExpansion
{\parbox[b]{2.9905in}{\begin{center}
\includegraphics[
height=1.849in,
width=2.9905in
]%
{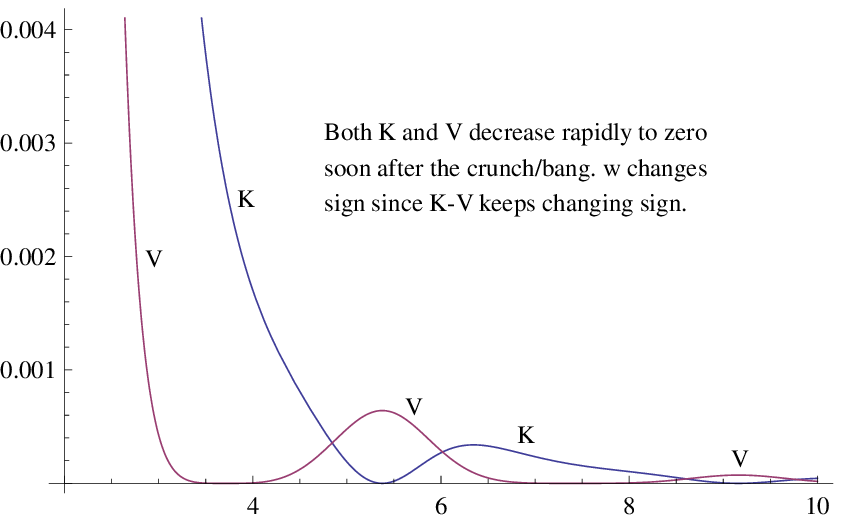}%
\\
FiG.20: The energy of $\sigma$ field decreases.
\end{center}}}%
%EndExpansion
&
%TCIMACRO{\FRAME{itbpFU}{3.0113in}{1.8853in}{0in}{\Qcb{FIG.21: Equation of
%state $w\left(  \tau\right)  .$}}{}{cyclicfig.w_{n}5b0.eps}%
%width 3.0113in;  height 1.8853in;  depth 0in;  original-width 3.333in;
%original-height 2.0781in;  cropleft "0";  croptop "1";  cropright "1";
%filename 'cyclicFig.w_n5b0.eps';file-properties "XNPEU";}}}%
%BeginExpansion
{\parbox[b]{3.0113in}{\begin{center}
\includegraphics[
height=1.8853in,
width=3.0113in
]%
{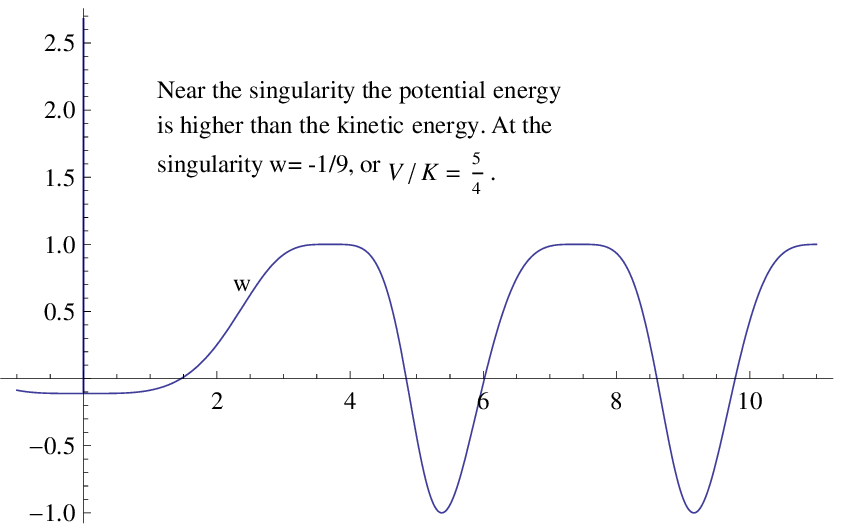}%
\\
FIG.21: Equation of state $w\left(  \tau\right)  .$
\end{center}}}%
%EndExpansion
\end{array}
\end{equation}
The behavior of the energy, pressure and the equation of state are indicated
on the figures.

The behavior of various quantities near the bang/crunch singularity is given
by the Taylor expansion near $\tau\rightarrow0$ as follows. These agree with
the $n=\infty$ limit of the previous cases%

\begin{align}
a\left(  \tau\right)   &  \rightarrow\frac{\kappa}{T\sqrt{48c}}\left(
\frac{\tau}{T}\right)  ^{3}\frac{1}{2\sqrt{5}}\left[  1-\frac{1}{60}\left(
\frac{\tau}{T}\right)  ^{4}+O\left(  \frac{\tau}{T}\right)  ^{8}\right] \\
\frac{\dot{a}\left(  \tau\right)  }{a\left(  \tau\right)  }  &  \rightarrow
\frac{1}{T}\left(  \frac{\tau}{T}\right)  ^{-1}\left[  3-\frac{1}{15}\left(
\frac{\tau}{T}\right)  ^{4}+O\left(  \frac{\tau}{T}\right)  ^{8}\right] \\
H\left(  \tau\right)   &  \rightarrow\frac{\sqrt{48c}}{\kappa}\left(
\frac{\tau}{T}\right)  ^{-4}\left[  6\sqrt{5}1-\frac{1}{6\sqrt{5}}\left(
\frac{\tau}{T}\right)  ^{4}+O\left(  \frac{\tau}{T}\right)  ^{8}\right] \\
\sigma\left(  \tau\right)   &  \rightarrow\frac{\sqrt{6}}{\kappa}\left[
-\ln\left(  \frac{\tau}{T}\right)  ^{2}+\frac{1}{2}\ln80+\frac{1}{240}\left(
\frac{\tau}{T}\right)  ^{4}+O\left(  \frac{\tau}{T}\right)  ^{8}\right] \\
\dot{\sigma}\left(  \tau\right)   &  \rightarrow\frac{\sqrt{6}}{\kappa
T}\left(  \frac{\tau}{T}\right)  ^{-1}\left[  -2+\frac{1}{60}\left(
\frac{\tau}{T}\right)  ^{4}+O\left(  \frac{\tau}{T}\right)  ^{8}\right] \\
V\left(  \sigma\left(  \tau\right)  \right)   &  \rightarrow\frac{288c}%
{\kappa^{4}}\left(  \frac{\tau}{T}\right)  ^{-8}\left[  50-\frac{5}{3}\left(
\frac{\tau}{T}\right)  ^{4}+O\left(  \frac{\tau}{T}\right)  ^{8}\right] \\
K\left(  \sigma\left(  \tau\right)  \right)   &  \rightarrow\frac{288c}%
{\kappa^{4}}\left(  \frac{\tau}{T}\right)  ^{-8}\left[  40+\frac{2}{3}\left(
\frac{\tau}{T}\right)  ^{4}+O\left(  \frac{\tau}{T}\right)  ^{8}\right] \\
w\left(  \tau\right)   &  \rightarrow-\frac{1}{9}+\frac{2}{81}\left(
\frac{\tau}{T}\right)  ^{4}+O\left(  \frac{\tau}{T}\right)  ^{8}\nonumber
\end{align}

\newpage

\section{The closed ($k=+1$) FRW universe \label{k+section}}

For $k=+1$ the system of equations is Eqs.(\ref{f}-\ref{E}). This amounts to
the motion of two particles $\phi,s$ satisfying the equations of motion
derived from Hamiltonians
\begin{equation}
H\left(  \phi\right)  =\frac{1}{2}\dot{\phi}^{2}+V_{b}\left(  \phi\right)
\text{,\ \ }H\left(  s\right)  =\frac{1}{2}\dot{s}^{2}+V_{c}\left(  s\right)
, \label{HsHf}%
\end{equation}
with
\begin{equation}
V_{c}\left(  s\right)  =\frac{1}{2}Ks^{2}+cs^{4},\;\text{and }V_{b}\left(
\phi\right)  =\frac{1}{2}K\phi^{2}-b\phi^{4}, \label{VsVf}%
\end{equation}
as plotted in Fig.(\ref{positive-k}) for $K=+1/r_{0}^{2}$, and whose energies
are constrained by%
\begin{equation}
H_{\phi}=H_{s}. \label{HsHf2}%
\end{equation}
The motion changes character depending on whether $E_{\phi}=E_{s}$ is larger
or smaller than the peak of the inverted double well in Fig.(\ref{positive-k}%
), i.e. the maximum of $V_{b>0}\left(  \phi\right)  $. This critical value is
given by
\begin{equation}
E^{\ast}=\frac{K^{2}}{16b}. \label{crit}%
\end{equation}
Therefore we need to discuss separately the high and low energy levels
$E_{\phi}=E_{s}$ above and below this critical value as shown in
Fig.(\ref{positive-k}).

\setcounter {figure} {21}%
%TCIMACRO{\FRAME{fhFU}{3.3797in}{2.1015in}{0pt}{\Qcb{The closed FRW universe,
%$k>0.$}}{\Qlb{positive-k}}{cyclicpotentialvk+.eps}%
%width 3.3797in;  height 2.1015in;  depth 0pt;  original-width 3.333in;
%original-height 2.0626in;  cropleft "0";  croptop "1";  cropright "1";
%filename 'cyclicPotentialVk+.eps';file-properties "XNPEU";}}}%
%BeginExpansion
\begin{figure}
[h]
\begin{center}
\includegraphics[
height=2.1015in,
width=3.3797in
]%
{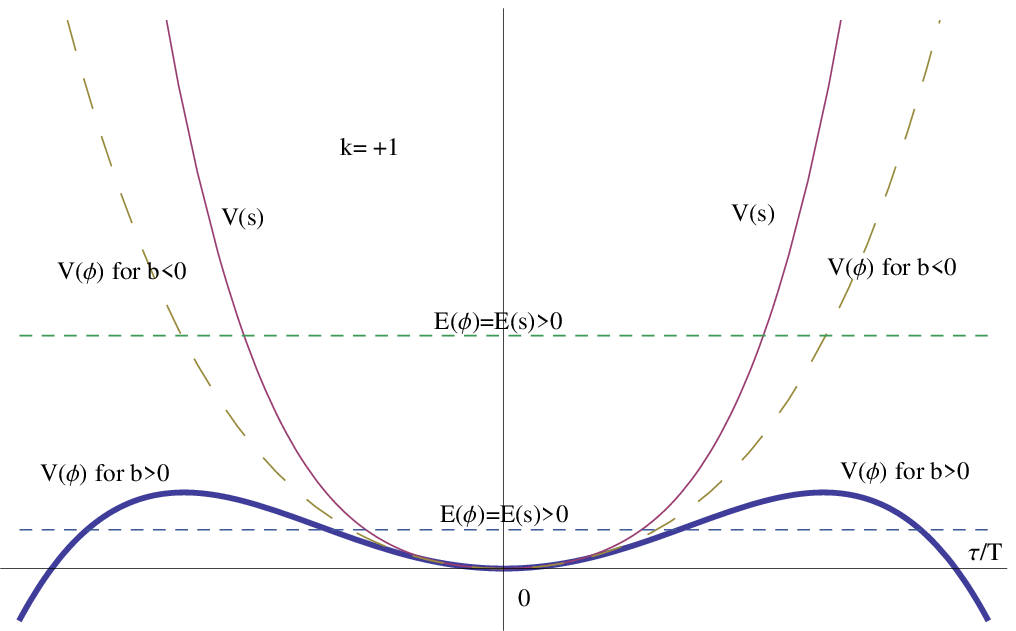}%
\caption{The closed FRW universe, $k>0.$}%
\label{positive-k}%
\end{center}
\end{figure}
%EndExpansion

\subsection{Higher level $E>E^{\ast},$ and $b>0$ or $b<0$ \label{highb+-}}

For the \textit{higher} level of $E_{s}=E_{\phi}>E^{\ast}$, the intuitive
physics discussion works in exactly the same way as the discussion for the
$k=0$ case at the beginning of section (\ref{k0section}), for both $b\geq0$ or
$b\leq0.$ In the case of $b>0,$ the particle $s_{\gamma}\left(  \tau\right)  $
is trapped in an infinite well and oscillates between turning points $\pm
s_{0}\left(  E\right)  $, while $\phi_{\gamma}\left(  \tau\right)  $
oscillates from minus infinity to plus infinity. In the case of $b<0$ both
particles are trapped in infinite wells, so they oscillate between turning
points $\pm\phi_{0}\left(  E\right)  $ and $\pm s_{0}\left(  E\right)  $
respectively. The turning points $\pm\phi_{0}\left(  E\right)  $, $\pm
s_{0}\left(  E\right)  $ are the points where the curves $V_{c}\left(
s\right)  ,V_{b<0}\left(  \phi\right)  $ intersect the horizontal curve
$E_{\phi}=E_{s}=E$ as seen in the figure.

Hence for all cases $b\geq0$ or $b\leq0$ at the \textit{higher} energy level
of $E_{s}=E_{\phi},$ the \textit{geodesically complete} motion is described by
plots of $\phi_{\gamma}\left(  \tau\right)  ,s_{\gamma}\left(  \tau\right)  $
that are similar in character to the $k=0$ case given in Figs.(\ref{fs5}-21).
In particular for very small curvature $K$ (large values of the curvature
radius $r_{0}$), the $K\neq0$ plots should approach the $K=0$ plots.
Therefore, we will not include the $K\neq0$ plots here.

The exact solutions for the \textit{higher} energy level are denoted as
$s_{\gamma}^{+}\left(  \tau\right)  ,\phi_{\gamma}^{+}\left(  \tau\right)  $
where the superscript \textquotedblleft$+$\textquotedblright\ refers to
$E>E^{\ast}$ with $E_{s}=E_{\phi}=E.$ The solution $s_{\gamma}^{+}\left(
\tau\right)  $ is given by
\begin{equation}
s_{\gamma}^{+}\left(  \tau\right)  =\sqrt{\frac{1-K^{2}T_{s}^{4}}{8cT_{s}^{2}%
}}\frac{sn\left(  \frac{\tau}{T_{s}}|m_{s}\right)  }{dn\left(  \frac{\tau
}{T_{s}}|m_{s}\right)  },\;\;%
\begin{array}
[c]{l}%
m_{s}\left(  E\right)  \equiv\frac{1}{2}\left(  1-KT_{s}^{2}\left(  E\right)
\right) \\
T_{s}\left(  E\right)  \equiv\left(  16cE+K^{2}\right)  ^{-1/4}%
\end{array}
\label{sk1}%
\end{equation}
while $\phi_{\gamma}^{+}\left(  \tau\right)  $ has the following expressions
for $b>0$%

\begin{equation}
\phi_{\gamma,b\geq0}^{+}\left(  \tau\right)  =\sqrt{\frac{1}{8bT_{+}^{2}}%
}\frac{sn\left(  \frac{\tau}{T_{+}}|m_{+}\right)  }{1+cn\left(  \frac{\tau
}{T_{+}}|m_{+}\right)  },~%
\begin{array}
[c]{l}%
m_{+}\left(  E\right)  \equiv\frac{1}{2}+KT_{+}^{2}\left(  E\right) \\
T_{+}\left(  E\right)  \equiv\left(  64bE\right)  ^{-1/4}%
\end{array}
,\; \label{fk1}%
\end{equation}
or $b<0$%

\begin{equation}
\phi_{\gamma,b\leq0}^{+}\left(  \tau\right)  =\sqrt{\frac{1-K^{2}T_{-}^{4}%
}{8\left\vert b\right\vert T_{-}^{2}}}\frac{sn\left(  \frac{\tau}{T_{-}}%
|m_{-}\right)  }{dn\left(  \frac{\tau}{T_{-}}|m_{-}\right)  },\text{ \ }%
\begin{array}
[c]{l}%
m_{-}\left(  E\right)  \equiv\frac{1}{2}\left(  1-KT_{-}^{2}\left(  E\right)
\right) \\
T_{-}\left(  E\right)  \equiv\left(  16\left\vert b\right\vert E+K^{2}\right)
^{-1/4}%
\end{array}
\label{fk2}%
\end{equation}
Note that all symbols $T,m$ that appear in the Jacobi elliptic functions are
determined in terms of the energy level $E=E_{s}=E_{\phi}$ as given above. The
$T_{\pm}\left(  E\right)  $ are determined in terms of $E,b,c,K$ ~by the
energy condition $E_{\phi}=E_{s}=E,$ by using the following expressions
computed from the Hamiltonians $H\left(  \phi\right)  ,H\left(  s\right)  $
given in Eqs.(\ref{HsHf}-\ref{HsHf2}) for the solutions $s_{\gamma}^{+}\left(
\tau\right)  ,\phi_{\gamma}^{+}\left(  \tau\right)  $ above
\begin{equation}
E_{s}=\frac{1}{16cT_{s}^{4}}\left(  1-K^{2}T_{s}^{4}\right)  ,\;E_{\phi}%
^{+}=\frac{1}{64bT_{+}^{4}},\;E_{\phi}^{-}=\frac{1}{16\left\vert b\right\vert
T_{-}^{4}}\left(  1-K^{2}T_{-}^{4}\right)  .\;
\end{equation}
Note also we have assumed that $E_{s}=E_{\phi}^{\pm}=E$ is higher than the
critical value $E^{\ast}$ in Eq.(\ref{crit}). This yields the expressions for
$T_{s,\pm}\left(  E\right)  ,m_{s,\pm}\left(  E\right)  $ given above as well
as the ranges for these parameters as a function of the energy level as
follows%
\begin{equation}
KT_{s}^{2}\left(  E\right)  <\sqrt{\frac{\left\vert b\right\vert }{\left\vert
b\right\vert +c}},\;KT_{-}^{2}\left(  E\right)  <\frac{1}{\sqrt{2}}%
,\;,~KT_{+}^{2}\left(  E\right)  <\frac{1}{2},
\end{equation}
which determines the possible range of values for $m_{s},m_{\pm}$ as $E$
changes in the range $E\geq E^{\ast}$
\begin{equation}
\frac{1}{2}\left(  1-\left(  1+c/\left\vert b\right\vert \right)
^{-1/2}\right)  <m_{s}\left(  E\right)  \leq\frac{1}{2};\;\;\;\;(\frac{1}%
{2}-\frac{\sqrt{2}}{4})<m_{-}\left(  E\right)  \leq\frac{1}{2};\;\;\frac{1}%
{2}\leq m_{+}\left(  E\right)  <1.
\end{equation}
It is easy to see that in the zero curvature limit $K\rightarrow0$ these
solutions reduce to Eqs.(\ref{fszerok}) for $b\geq0$ or Eqs.(\ref{b-zerok})
for $b\leq0.$

As they stand these solutions do not yet satisfy the requirement $\phi
^{2}\left(  \tau\right)  \geq s^{2}\left(  \tau\right)  $ at all times. This
can be satisfied only by requiring the period of $\phi$ to be a multiple
integer of the period of $s.$ The analytic expression for this conditions is
\begin{align}
b  &  \geq0:\;T_{+}Q\left(  m_{+}\right)  =2nT_{s}Q\left(  m_{s}\right)
,\;n=1,2,3,\cdots\label{quantum+}\\
b  &  \leq0:\;T_{-}Q\left(  m_{-}\right)  =nT_{s}Q\left(  m_{s}\right)
,\;n=1,2,3,\cdots\label{quantum-}%
\end{align}
where the quantity $Q\left(  z\right)  $ is a well known special function,
namely the quarter period of the corresponding Jacobi elliptic function, and
is given by the following integral representation of the EllipticK integral%
\begin{equation}
Q\left(  z\right)  =\int_{0}^{\pi/2}\frac{d\theta}{\sqrt{1-z\sin^{2}\theta}%
}=\text{EllipticK}\left(  z\right)  .
\end{equation}
The consequence of this is to require a certain combination of parameters
$\left(  b,c,K,E\right)  $ to be quantized. The range of parameters in the
model $\left(  b,c,K,E\right)  $ that can satisfy the constraint can be
determined numerically\footnote{For example, using Mathematica, which
recognizes the function EllipticK(z), one can plot $T_{\pm}Q\left(  m_{\pm
}\right)  /T_{s}Q\left(  m_{s}\right)  $ as a function of one of the
parameters $E,b,c,K$ (example $b$) while the other three are chosen
arbitrarily. When the plot matches an integer $n$, this fixes the value of the
remaining parameter (i.e. $b$ in the example above) in terms of the integer
$n$ and the other three parameters. \label{numerical}} by using the
expressions above. Therefore only a model that can satisfy this condition can
give the corresponding geodesically complete solutions.

\subsection{Lower level $E<E^{\ast},$ and $b<0$}

When the energy is less than the critical value, the exact solutions are
denoted as $s_{\gamma}^{-}\left(  \tau\right)  ,\phi_{\gamma}^{-}\left(
\tau\right)  $ where the superscript \textquotedblleft$-$\textquotedblright%
\ refers to the energy interval $0<E_{s}=E_{\phi}=E<E^{\ast}.$ We consider at
first the case of $b<0,$ for which $V_{b<0}\left(  \phi\right)  $ is
represented by the dashed curve. There is no difference in this case with the
$b<0$ case above, so the formulas above apply, namely $s_{\gamma}^{-}\left(
\tau\right)  $ is the same as $s_{\gamma}^{+}\left(  \tau\right)  $, and
$\phi_{\gamma,b<0}^{-}\left(  \tau\right)  $ is the same as $\phi_{\gamma
,b<0}^{+}\left(  \tau\right)  ,$ except for the fact that now the energy is in
the range $0<E<E^{\ast}$
\begin{equation}
s_{\gamma,b<0}^{-}\left(  \tau\right)  =\sqrt{\frac{1-K^{2}T_{s}^{4}}%
{8cT_{s}^{2}}}\frac{sn\left(  \frac{\tau}{T_{s}}|m_{s}\right)  }{dn\left(
\frac{\tau}{T_{s}}|m_{s}\right)  },\;\;%
\begin{array}
[c]{l}%
m_{s}\left(  E\right)  \equiv\frac{1}{2}\left(  1-KT_{s}^{2}\left(  E\right)
\right) \\
T_{s}\left(  E\right)  \equiv\left(  16cE+K^{2}\right)  ^{-1/4}%
\end{array}
, \label{s-}%
\end{equation}%
\begin{equation}
\phi_{\gamma,b<0}^{-}\left(  \tau\right)  =\sqrt{\frac{1-K^{2}T_{-}^{4}%
}{8\left\vert b\right\vert T_{-}^{2}}}\frac{sn\left(  \frac{\tau}{T_{-}}%
|m_{-}\right)  }{dn\left(  \frac{\tau}{T_{-}}|m_{-}\right)  },\text{ }%
\begin{array}
[c]{l}%
m_{-}\left(  E\right)  \equiv\frac{1}{2}\left(  1-KT_{-}^{2}\left(  E\right)
\right) \\
T_{-}\left(  E\right)  \equiv\left(  16\left\vert b\right\vert E+K^{2}\right)
^{-1/4}%
\end{array}
.
\end{equation}
Since the energy is less than the critical value, $0\leq E_{s}=E_{\phi}\leq
E^{\ast},$ we must now restrict the range of the parameters $m,T$ to
\begin{equation}
\sqrt{\frac{\left\vert b\right\vert }{\left\vert b\right\vert +c}}\leq
KT_{s}^{2}\left(  E\right)  \leq1,\;\frac{1}{\sqrt{2}}\leq KT_{-}^{2}\left(
E\right)  \leq1,
\end{equation}
which implies%
\begin{equation}
0\leq m_{s}\left(  E\right)  \leq\frac{1}{2}\left(  1-\sqrt{\frac{\left\vert
b\right\vert }{\left\vert b\right\vert +c}}\right)  ,\;m_{-}\left(  E\right)
\leq\frac{1}{2}\left(  1-\frac{1}{\sqrt{2}}\right)  .
\end{equation}
To obtain geodesically complete solutions the analog of the quantization
condition in Eq.(\ref{quantum-}) must be further imposed, $T_{-}Q\left(
m_{-}\right)  =nTQ\left(  m\right)  $.

\subsection{Lower level $E<E^{\ast},$ and $b>0,$ finite bounce \label{bounceB}%
}

For the case of $b>0$ ($V_{b>0}\left(  \phi\right)  $ represented by the
inverted double well in Fig.(\ref{positive-k})) there are new features. For
$\phi$ now there is the possibility to either be trapped inside the false
vacuum, or be outside of it, depending on initial conditions. When $\phi$ is
trapped in the false vacuum it should oscillate between two turning points;
when it is outside it should oscillate between a finite value and infinity. So
the solution has the form%
\begin{align}
\phi_{\gamma,b\geq0}^{-,in}\left(  \tau\right)   &  =\frac{\sqrt{KT_{in}%
^{2}-1}}{\sqrt{2b}T_{in}}sn\left(  \frac{\tau}{T_{in}}|m_{in}\right)
,\;\text{ }%
\begin{array}
[c]{l}%
m_{in}\left(  E\right)  \equiv\left(  KT_{in}^{2}\left(  E\right)  -1\right)
\\
T_{in}\left(  E\right)  \equiv\left(  \frac{K}{2}+\sqrt{\frac{K^{2}}{4}%
-4bE}\right)  ^{-1/2}%
\end{array}
,\label{finside}\\
\phi_{\gamma,b\geq0}^{-,out}\left(  \tau\right)   &  =\frac{\sqrt{KT_{out}%
^{2}+1}}{\sqrt{2b}T_{out}}\frac{1}{cn\left(  \frac{\tau+\tau_{0}}{T_{out}%
}|m_{out}\right)  },~%
\begin{array}
[c]{l}%
m_{out}\left(  E\right)  \equiv-\frac{1}{2}\left(  KT_{out}^{2}\left(
E\right)  -1\right) \\
T_{out}\left(  E\right)  \equiv\left(  K^{2}-16bE\right)  ^{-1/4}%
\end{array}
. \label{bounce1f}%
\end{align}
Meanwhile $s\left(  \tau\right)  $ continues to oscillate as before between
two turning points, so it is still given by the same expression, namely
\begin{equation}
s_{\gamma,b>0}^{-}\left(  \tau\right)  =\sqrt{\frac{1-K^{2}T_{s}^{4}}%
{8cT_{s}^{2}}}\frac{sn\left(  \frac{\tau}{T_{s}}|m_{s}\right)  }{dn\left(
\frac{\tau}{T_{s}}|m_{s}\right)  },\;\;%
\begin{array}
[c]{l}%
m_{s}\left(  E\right)  \equiv\frac{1}{2}\left(  1-KT_{s}^{2}\left(  E\right)
\right) \\
T_{s}\left(  E\right)  \equiv\left(  16cE+K^{2}\right)  ^{-1/4}%
\end{array}
. \label{bounce1s}%
\end{equation}
The energies of these solutions are computed in terms of the parameters $m,T$
by using the Hamiltonians in Eqs.(\ref{HsHf}-\ref{HsHf2}) as follows
\begin{equation}
E_{s}=\frac{1}{16cT_{s}^{4}}\left(  1-K^{2}T_{s}^{4}\right)  ,\;E_{\phi}%
^{in}=\frac{KT_{in}^{2}-1}{4bT_{in}^{4}},\;E_{\phi}^{out}=\frac{K^{2}%
T_{out}^{4}-1}{16bT_{out}^{4}}.
\end{equation}
All energies must be positive and equal to each other $E_{\phi}^{out}=E_{s}=E$
and $E_{\phi}^{in}=E_{s}=E,$ as well as smaller than $E^{\ast}.$ This yields
the expressions for $T_{s,in,out}\left(  E\right)  $ and $m_{s,in,out}\left(
E\right)  $ given above as well as the ranges for these parameters as a
function of $E$, as follows$\allowbreak\allowbreak\allowbreak$%
\[
\;1\geq KT_{s}^{2}\left(  E\right)  \geq\sqrt{\frac{b}{c+b}},\;\;2\geq
KT_{in}^{2}\left(  E\right)  \geq1,\;KT_{out}^{2}\left(  E\right)  \geq1.
\]
Similarly, the range of values for the parameters $m_{s},m_{in},m_{out}$ are
then as\allowbreak\ follows\allowbreak%
\begin{equation}
0\leq m_{s}\left(  E\right)  \leq\frac{1}{2}\left(  1-\sqrt{\frac{b}{b+c}%
}\right)  ,\;0\leq m_{in}\left(  E\right)  \leq1,\;m_{out}\left(  E\right)
\leq0.
\end{equation}

To obtain the geodesically complete solution, in the case of the
\textit{inside} solution the quantization condition $T_{in}Q\left(
m_{in}\right)  =nTQ\left(  m\right)  $ is required$^{\text{\ref{numerical}}}$.
However, in the case of the $outside$ solution no quantization condition is
needed as explained below.

\begin{figure}
[h]
\begin{center}
\includegraphics[
height=2.38in,
width=3.8674in
]%
{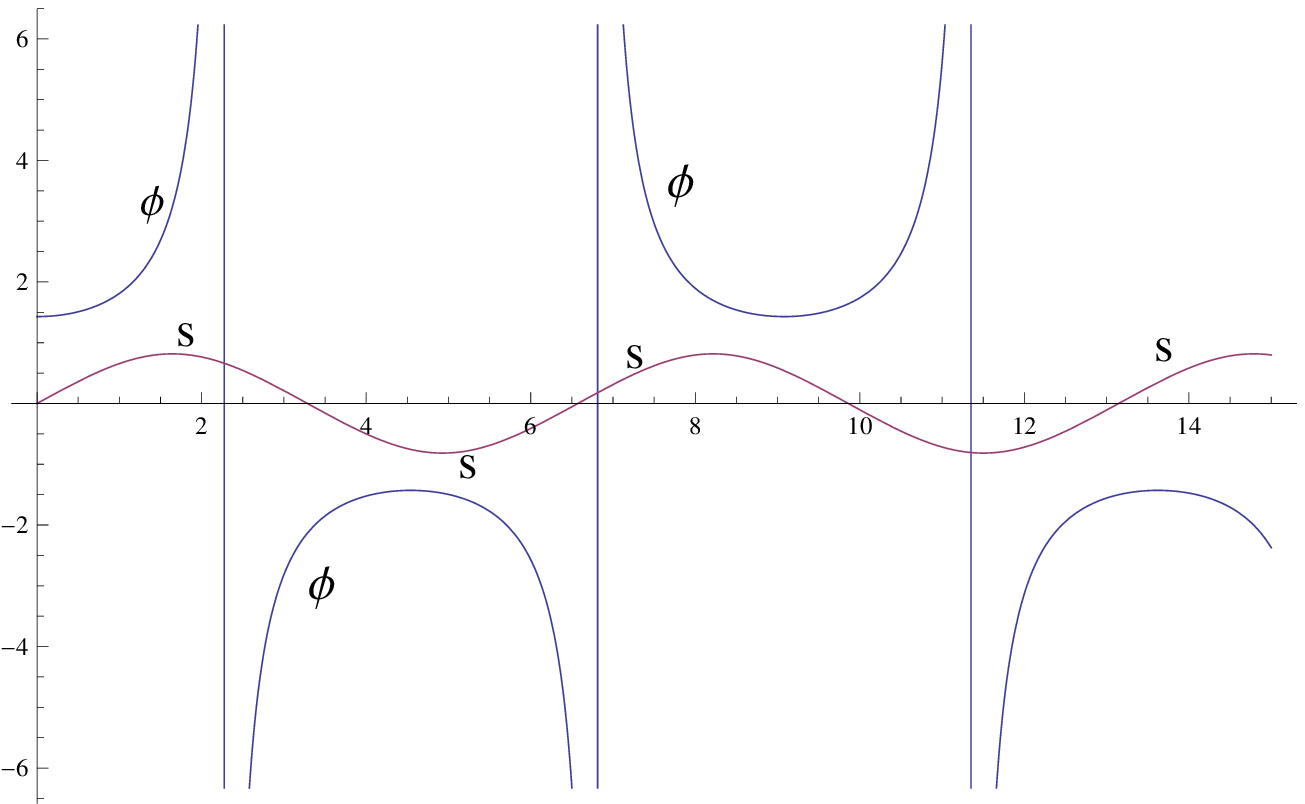}%
\caption{$\phi_{\gamma,b>0}^{-,out}\left(  \tau\right)  $ and $s_{\gamma
,b>0}^{-}\left(  \tau\right)  $ for the bouncing solution.}%
\label{bounce1}%
\end{center}
\end{figure}

We now comment on the outside solution given by $\phi_{\gamma,b>0}%
^{-,out}\left(  \tau\right)  ,s_{\gamma,b>0}^{-}\left(  \tau\right)  $ in
Eqs.(\ref{bounce1f},\ref{bounce1s}) because it is different in character as
compared to all the previous cases. It describes a periodically
contracting/expanding universe with bounces that occur at minimum finite
values of the scale factor, while the maximum is infinite. $\phi_{\gamma
,b>0}^{-,out}\left(  \tau\right)  $ describes the behavior of $\phi$ outside
of the false vacuum, the amplitude for $\phi$ is always larger than the
amplitude for $s_{\gamma,b>0}^{-}\left(  \tau\right)  $ at all times and for
all boundary conditions, including the additional arbitrary parameter
$\tau_{0}$. This solution represents \textit{cyclic bounces at finite minimum
sizes of the universe}. This is shown in Fig.(\ref{bounce1}).
%TCIMACRO{\FRAME{fhFU}{3.8674in}{2.38in}{0pt}{\Qcb{$\phi_{\gamma,b>0}%
%^{-,out}\left(  \tau\right)  $ and $s_{\gamma,b>0}^{-}\left(  \tau\right)  $
%for the bouncing solution.}}{\Qlb{bounce1}}{cyclicfig.bounce.eps}%
%width 3.8674in;  height 2.38in;  depth 0pt;  original-width 5.1923in;
%original-height 3.186in;  cropleft "0";  croptop "1";  cropright "1";
%filename 'cyclicFig.bounce.eps';file-properties "XNPEU";}}}%
%BeginExpansion
%EndExpansion
The cyclic bounce occurs for all values of $c>0$, all values of $b>0,$ and all
values of the relative initial conditions $\tau_{0}$ at $\tau=0$. This is why
we added an additional phase $\tau_{0}$ in the expression of $\phi
_{\gamma,b>0}^{-,out}\left(  \tau\right)  $. There is no need to synchronize
the boundary conditions at $\tau=0$ for $\phi\left(  0\right)  ,s\left(
0\right)  $ in order to get geodesically complete solutions that satisfy
$\phi^{2}\left(  \tau\right)  \geq s^{2}\left(  \tau\right)  $ at all times.
Moreover, $b/c$ need not be quantized since the periods of $\phi,s$ can now be
arbitrary relative to each other.

\begin{figure}
[h]
\begin{center}
\includegraphics[
height=2.1015in,
width=3.3797in
]%
{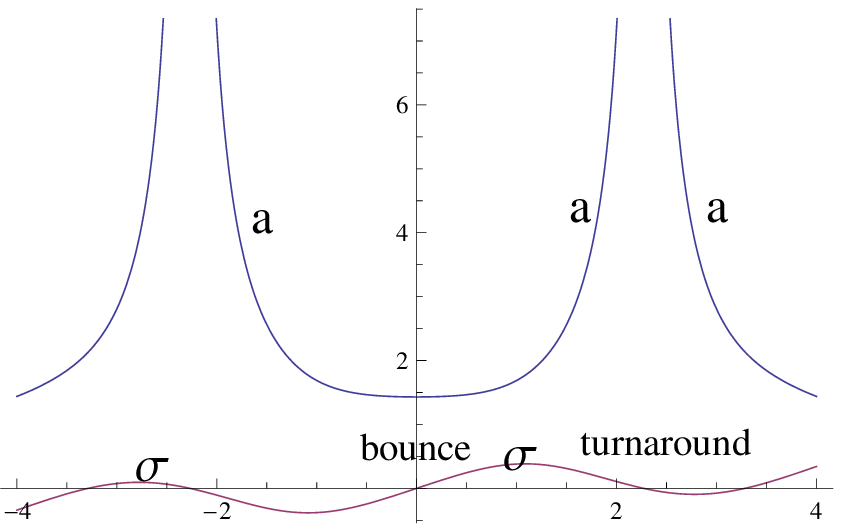}%
\caption{The bounce and turnaround.}%
\label{bounce_a+sig}%
\end{center}
\end{figure}

The corresponding plots for the scale factor $a\left(  \tau\right)  $ and
$\sigma\left(  \tau\right)  $ for the bounce are given in
Fig.(\ref{bounce_a+sig}).
%TCIMACRO{\FRAME{fhFU}{3.3797in}{2.1015in}{0pt}{\Qcb{The bounce and
%turnaround.}}{\Qlb{bounce_a+sig}}{cyclicfig.bounce_a+sig.eps}%
%width 3.3797in;  height 2.1015in;  depth 0pt;  original-width 3.333in;
%original-height 2.0626in;  cropleft "0";  croptop "1";  cropright "1";
%filename 'cyclicFig.bounce_a+sig.eps';file-properties "XNPEU";}}}%
%BeginExpansion
%EndExpansion
The minimum size of the scale factor $a\left(  \tau\right)  \sim\sqrt
{\phi_{\gamma}^{2}\left(  \tau\right)  -s_{\gamma}^{2}\left(  \tau\right)  }$
at the bounce instant is not the same each time since the initial values of
$\phi\left(  \tau\right)  ,s\left(  \tau\right)  $ are not synchronized and
their periods are not related.

For this bounce solution to play a physical role in cosmology we need the
curvature terms $K\phi^{2},Ks^{2}$ to be able to compete with the potential
terms $b\phi^{4},cs^{4}$. For this to be phenomenologically tenable in a
complete cosmological model, a period of inflation must follow after the bang
so that the universe inflates to its current size and to its almost flat
current condition (since $r_{0}$ would not be identified with today's size of
the universe).

As a limiting case of the above solutions we point out the special case of the
$b>0$ scenario when $E=E^{\ast}.$ As seen from Fig.(\ref{positive-k}), the
$\phi$ field sits still on top of the hill while the $s$ field oscillates in a
finite range. The maximum size of the universe is a finite number determined
by the constant value of the $\phi$ field.

\subsection{The case of $c<0$ \label{bounce2}}

There are also geodesically complete solutions when $c<0$ and $b>0$ which we
will outline very briefly. The $\phi,s$ Hamiltonians are the same as before,
as in Eq.(\ref{HsHf}-\ref{HsHf2}), with $K>0.$ There exist two interesting
classes of geodesically complete solutions that satisfy $\phi^{2}-s^{2}\geq0,$
which can occur when $c\,<0$ and $b>0.$ This happens when the corresponding
potentials $V\left(  \phi\right)  ,V\left(  s\right)  $ take the form in Figs.(25,26).

\begin{center}%
%TCIMACRO{\FRAME{itbpFU}{2.872in}{1.7867in}{0in}{\Qcb{FIG. 25 : $s$ oscillates
%inside, $\phi$ outside.}}{\Qlb{Cminus2}}{cyclicfig.cminus2.eps}%
%{\special{ language "Scientific Word";  type "GRAPHIC";  display "USEDEF";
%valid_file "F";  width 2.872in;  height 1.7867in;  depth 0in;
%original-width 3.333in;  original-height 2.0626in;  cropleft "0";
%filename 'CyclicFig.Cminus2.eps';file-properties "XNPEU";}}}%
%BeginExpansion
{\parbox[b]{2.872in}{\begin{center}
\includegraphics[
height=1.7867in,
width=2.872in
]%
{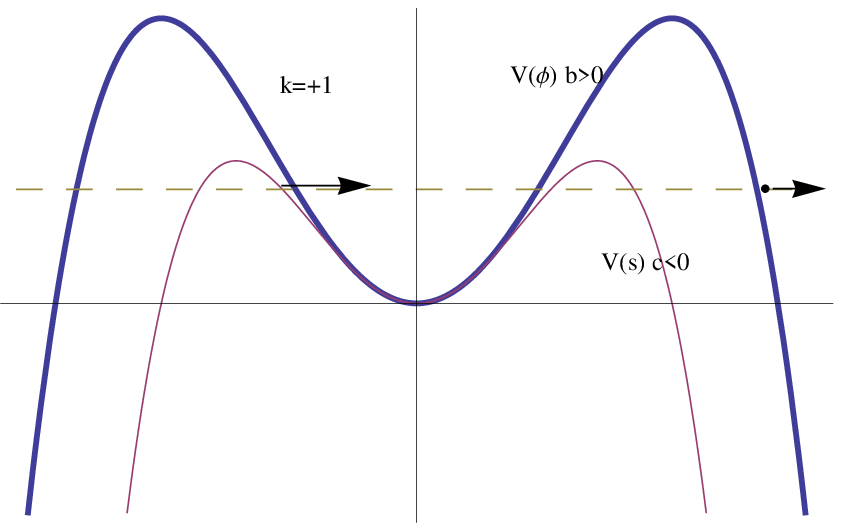}%
\\
FIG. 25 : $s$ oscillates inside, $\phi$ outside.
\end{center}}}%
%EndExpansion%
%TCIMACRO{\FRAME{ihFU}{2.8167in}{1.7521in}{0in}{\Qcb{FIG.26 : $s$ inside,
%$\phi$ depends on $E$.}}{\Qlb{Cminus1}}{cyclicfig.cminus1.eps}%
%width 2.8167in;  height 1.7521in;  depth 0in;  original-width 3.333in;
%original-height 2.0626in;  cropleft "0";  croptop "1";  cropright "1";
%filename 'CyclicFig.Cminus1.eps';file-properties "XNPEU";}}}%
%BeginExpansion
{\parbox[b]{2.8167in}{\begin{center}
\includegraphics[
height=1.7521in,
width=2.8167in
]%
{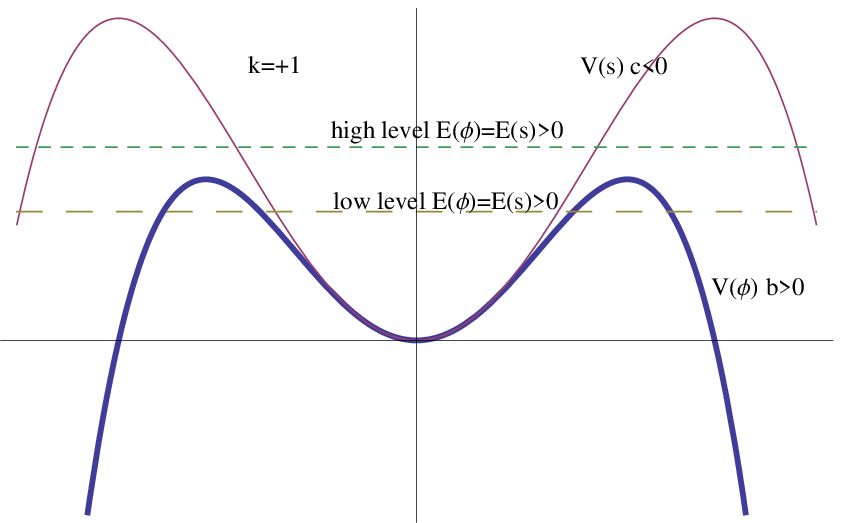}%
\\
FIG.26 : $s$ inside, $\phi$ depends on $E$.
\end{center}}}%
%EndExpansion
\ \
\end{center}

The first case is depicted in Fig.(25) when $0\leq E\leq V_{\max}\left(
s\right)  =\frac{K^{2}}{16\left\vert c\right\vert },$ and $V_{\max}\left(
\phi\right)  =\frac{K^{2}}{16b}$ is higher, i.e. with $b<\left\vert
c\right\vert .$ Then $s$ oscillates in the region of a false vacuum, while
$\phi$ oscillates outside of the false vacuum all the way to infinity. This is
similar to the finite bounce solution we discussed in section (\ref{bounceB})
and figures (\ref{bounce1},\ref{bounce_a+sig}). For completeness we record the
solution
\begin{equation}
s_{\gamma,c\leq0}^{-}\left(  \tau\right)  =\frac{\sqrt{KT_{c}^{2}-1}}%
{\sqrt{2\left\vert c\right\vert }T_{c}}sn\left(  \frac{\tau}{T_{c}}%
|m_{c}\right)  ,\;%
\begin{array}
[c]{l}%
m_{c}\left(  E\right)  \equiv\left(  KT_{c}^{2}\left(  E\right)  -1\right) \\
T_{c}\left(  E\right)  \equiv\left(  \frac{K}{2}+\sqrt{\frac{K^{2}}%
{4}-4\left\vert c\right\vert E}\right)  ^{-1/2}%
\end{array}
,
\end{equation}
and%
\begin{equation}
\phi_{\gamma,b\geq0}^{-,out}\left(  \tau\right)  =\frac{\sqrt{KT_{out}^{2}+1}%
}{\sqrt{2b}T_{out}}\frac{1}{cn\left(  \frac{\tau+\tau_{0}}{T_{out}}%
|m_{out}\right)  },~%
\begin{array}
[c]{l}%
m_{out}\left(  E\right)  \equiv-\frac{1}{2}\left(  KT_{out}^{2}\left(
E\right)  -1\right) \\
T_{out}\left(  E\right)  \equiv\left(  K^{2}-16bE\right)  ^{-1/4}%
\end{array}
. \label{t02}%
\end{equation}
As in the previous case of the bounce, this solution also represents
\textit{cyclic bounces at finite minimum sizes of the universe}, similar to
those in Figs.(\ref{bounce1},\ref{bounce_a+sig}). It occurs for all values of
$b>0$ and $c<0$ provided $b<\left\vert c\right\vert ,$ and provided $K$ is
large enough so that the curvature terms $K\phi^{2},Ks^{2}$ are able to
compete with the potential term $b\phi^{4},cs^{4}.$ The parameters ($c,b,K$)
do not need to satisfy any quantization conditions. Also, the synchronization
of the relative phase is not needed, hence we have allowed the additional
integration constant $\tau_{0}$ in the solution in Eq.(\ref{t02}).

The second case is depicted in Fig.(26) when $0\leq E\leq V_{\max}\left(
s\right)  ,$ where $V_{\max}\left(  s\right)  =\frac{K^{2}}{16\left\vert
c\right\vert }$ and $V_{\max}\left(  \phi\right)  =\frac{K^{2}}{16b},$ with
$b>\left\vert c\right\vert .$ Then $s$ oscillates in the region of a false
vacuum, while the behavior of $\phi$ depends on whether the energy is low
(below $V_{\max}\left(  \phi\right)  $) or high (between $V_{\max}\left(
\phi\right)  $ and $V_{\max}\left(  s\right)  $). The analytic solutions are
similar to the ones discussed above, except that now $c<0,$ and the energy $E$
is limited to the region $0\leq E\leq V_{\max}\left(  s\right)  .$ For
completeness we list the geodesically complete solutions that satisfy
$\phi^{2}-s^{2}\geq0,$ together with the quantization condition for their
periods
\begin{equation}
\text{high }:\;\left\{
\begin{array}
[c]{l}%
s_{\gamma,c\leq0}^{+}\left(  \tau\right)  =\frac{\sqrt{KT_{c}^{2}-1}}%
{\sqrt{2\left\vert c\right\vert }T_{c}}sn\left(  \frac{\tau}{T_{c}}%
|m_{c}\right)  ,\;%
\begin{array}
[c]{l}%
m_{c}\left(  E\right)  \equiv\left(  KT_{c}^{2}\left(  E\right)  -1\right) \\
T_{c}\left(  E\right)  \equiv\left(  \frac{K}{2}+\sqrt{\frac{K^{2}}%
{4}-4\left\vert c\right\vert E}\right)  ^{-1/2}%
\end{array}
\\
\phi_{\gamma,b\geq0}^{+}\left(  \tau\right)  =\sqrt{\frac{1}{8bT_{+}^{2}}%
}\frac{sn\left(  \frac{\tau}{T_{+}}|m_{+}\right)  }{1+cn\left(  \frac{\tau
}{T_{+}}|m_{+}\right)  },~%
\begin{array}
[c]{l}%
m_{+}\left(  E\right)  \equiv\frac{1}{2}+KT_{+}^{2}\left(  E\right) \\
T_{+}\left(  E\right)  \equiv\left(  64bE\right)  ^{-1/4}%
\end{array}
\\
T_{+}Q\left(  m_{+}\right)  =2nT_{c}Q\left(  m_{c}\right)
\end{array}
\right.  ,
\end{equation}
and
\begin{equation}
\text{low~~:}\;\left\{
\begin{array}
[c]{l}%
s_{\gamma,c\leq0}^{+}\left(  \tau\right)  =\frac{\sqrt{KT_{c}^{2}-1}}%
{\sqrt{2\left\vert c\right\vert }T_{c}}sn\left(  \frac{\tau}{T_{c}}%
|m_{c}\right)  ,\;%
\begin{array}
[c]{l}%
m_{c}\left(  E\right)  \equiv\left(  KT_{c}^{2}\left(  E\right)  -1\right) \\
T_{c}\left(  E\right)  \equiv\left(  \frac{K}{2}+\sqrt{\frac{K^{2}}%
{4}-4\left\vert c\right\vert E}\right)  ^{-1/2}%
\end{array}
\\
\phi_{\gamma,b\geq0}^{+,in}\left(  \tau\right)  =\frac{\sqrt{KT_{b}^{2}-1}%
}{\sqrt{2\left\vert c\right\vert }T_{b}}sn\left(  \frac{\tau}{T_{b}}%
|m_{b}\right)  ,\;%
\begin{array}
[c]{l}%
m_{b}\left(  E\right)  \equiv\left(  KT_{b}^{2}\left(  E\right)  -1\right) \\
T_{b}\left(  E\right)  \equiv\left(  \frac{K}{2}+\sqrt{\frac{K^{2}}{4}%
-4bE}\right)  ^{-1/2}%
\end{array}
\\
T_{b}Q\left(  m_{b}\right)  =nT_{c}Q\left(  m_{c}\right)
\end{array}
\right.  .
\end{equation}
In the low energy level it is also possible for $\phi$ to oscillate on the
outside of the false vacuum. In that case the solution is given by
\begin{equation}
\text{low~~:}\;\left\{
\begin{array}
[c]{l}%
s_{\gamma,c\leq0}^{+}\left(  \tau\right)  =\frac{\sqrt{KT_{c}^{2}-1}}%
{\sqrt{2\left\vert c\right\vert }T_{c}}sn\left(  \frac{\tau}{T_{c}}%
|m_{c}\right)  ,\;%
\begin{array}
[c]{l}%
m_{c}\left(  E\right)  \equiv\left(  KT_{c}^{2}\left(  E\right)  -1\right) \\
T_{c}\left(  E\right)  \equiv\left(  \frac{K}{2}+\sqrt{\frac{K^{2}}%
{4}-4\left\vert c\right\vert E}\right)  ^{-1/2}%
\end{array}
,\\
\phi_{\gamma,b\geq0}^{+,out}\left(  \tau\right)  =\frac{\sqrt{KT_{out}^{2}+1}%
}{\sqrt{2b}T_{out}}\frac{1}{cn\left(  \frac{\tau+\tau_{0}}{T_{out}}%
|m_{out}\right)  },~%
\begin{array}
[c]{l}%
m_{out}\left(  E\right)  \equiv-\frac{1}{2}\left(  KT_{out}^{2}\left(
E\right)  -1\right) \\
T_{out}\left(  E\right)  \equiv\left(  K^{2}-16bE\right)  ^{-1/4}%
\end{array}
.
\end{array}
\right.
\end{equation}
In this case there is no quantization condition on the parameters, and there
is an additional integration constant $\tau_{0}$ which is arbitrary.

\section{The open ($k=-1$) FRW universe \label{k-section}}

For $k=-1,$ the dynamics of $\phi,s$ is described by the Hamiltonians in
Eqs.(\ref{HsHf},\ref{HsHf2}) with $K<0.$ The corresponding potentials
$V\left(  \phi\right)  ,V\left(  s\right)  $ are plotted in
Fig.(\ref{negative-k}). \setcounter {figure} {26}
%TCIMACRO{\FRAME{fhFU}{3.3797in}{2.1015in}{0pt}{\Qcb{The open FRW universe,
%$k<0.$}}{\Qlb{negative-k}}{cyclicpotentialvk-.eps}%
%width 3.3797in;  height 2.1015in;  depth 0pt;  original-width 3.333in;
%original-height 2.0626in;  cropleft "0";  croptop "1";  cropright "1";
%filename 'cyclicPotentialVk-.eps';file-properties "XNPEU";}}}%
%BeginExpansion
\begin{figure}
[h]
\begin{center}
\includegraphics[
height=2.1015in,
width=3.3797in
]%
{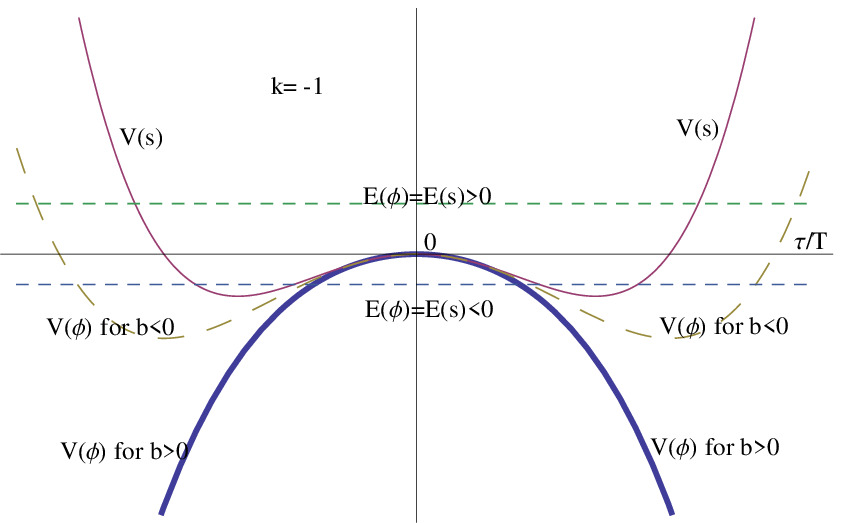}%
\caption{The open FRW universe, $k<0.$}%
\label{negative-k}%
\end{center}
\end{figure}
%EndExpansion

For the negative energy level $E_{s}=E_{\phi}<0,$ there cannot exist
geodesically complete solutions that satisfy $\phi^{2}\left(  \tau\right)
-s^{2}\left(  \tau\right)  \geq0$ at all times because of the following
argument. The energy $E_{s}=E_{\phi}$ must be above the minimum of $V\left(
s\right)  $ at the double well to have a solution for $s.$ Then the quantity
$\phi^{2}-s^{2}$ always changes sign, whether $b>0$ or $b<0$. For example,
suppose $s$ is at the minimum of $V\left(  s\right)  ,$ so the solution
$s\left(  \tau\right)  $ is just a constant. But $\phi\left(  \tau\right)  $
oscillates between two turning points $\phi_{\min}\left(  E\right)
<\phi\left(  \tau\right)  <\phi_{\max}\left(  E\right)  ,$ while sometimes
$\phi^{2}-s^{2}$ is positive and sometimes it is negative. The figure above is
drawn for the case $\left\vert b\right\vert <c.$ If we take $\left\vert
b\right\vert >c$ then $V\left(  s\right)  $ will be like the dashed curve and
while $V\left(  \phi\right)  $ will be like the solid thin curve (for $b<0$)
or the solid thick curve (for $b>0$). In these cases again there are no
solutions such that $\phi^{2}-s^{2}$ remains positive at all times.

For the positive energy level $E_{s}=E_{\phi}>0,$ there are geodesically
complete solutions that satisfy $\phi^{2}-s^{2}>0$ at all times. In fact this
case is formally identical to the case discussed in section (\ref{highb+-}).
In the present case $K<0,$ $c>0,$ while $b$ can have either sign. The exact
solutions are parallel to Eqs.(\ref{sk1}-\ref{fk2}) except for replacing
$K=-\left\vert K\right\vert .$ Thus, the solutions are
\begin{equation}
s_{\gamma}^{+}\left(  \tau\right)  =\sqrt{\frac{1-K^{2}T_{s}^{4}}{8cT_{s}^{2}%
}}\frac{sn\left(  \frac{\tau}{T_{s}}|m_{s}\right)  }{dn\left(  \frac{\tau
}{T_{s}}|m_{s}\right)  },\;\;%
\begin{array}
[c]{l}%
m_{s}\left(  E\right)  \equiv\frac{1}{2}\left(  1+\left\vert K\right\vert
T_{s}^{2}\left(  E\right)  \right) \\
T_{s}\left(  E\right)  \equiv\left(  16cE+K^{2}\right)  ^{-1/4}%
\end{array}
\end{equation}
while $\phi_{\gamma}^{+}\left(  \tau\right)  $ has the following expressions
for $b>0$%

\begin{equation}
\phi_{\gamma,b\geq0}^{+}\left(  \tau\right)  =\sqrt{\frac{1}{8bT_{+}^{2}}%
}\frac{sn\left(  \frac{\tau}{T_{+}}|m_{+}\right)  }{1+cn\left(  \frac{\tau
}{T_{+}}|m_{+}\right)  },~%
\begin{array}
[c]{l}%
m_{+}\left(  E\right)  \equiv\frac{1}{2}-\left\vert K\right\vert T_{+}%
^{2}\left(  E\right) \\
T_{+}\left(  E\right)  \equiv\left(  64bE\right)  ^{-1/4}%
\end{array}
,\;
\end{equation}
or $b<0$%

\begin{equation}
\phi_{\gamma,b\leq0}^{+}\left(  \tau\right)  =\sqrt{\frac{1-K^{2}T_{-}^{4}%
}{8\left\vert b\right\vert T_{-}^{2}}}\frac{sn\left(  \frac{\tau}{T_{-}}%
|m_{-}\right)  }{dn\left(  \frac{\tau}{T_{-}}|m_{-}\right)  },\text{ \ }%
\begin{array}
[c]{l}%
m_{-}\left(  E\right)  \equiv\frac{1}{2}\left(  1+\left\vert K\right\vert
T_{-}^{2}\left(  E\right)  \right) \\
T_{-}\left(  E\right)  \equiv\left(  16\left\vert b\right\vert E+K^{2}\right)
^{-1/4}%
\end{array}
\end{equation}

As they stand these solutions do not yet satisfy the requirement $\phi
^{2}\left(  \tau\right)  \geq s^{2}\left(  \tau\right)  $ at all times. This
can be satisfied only by requiring the period of $\phi$ to be a multiple
integer of the period of $s.$ The analytic expression for this conditions is,
as before
\begin{align}
b  &  \geq0:\;T_{+}Q\left(  m_{+}\right)  =2nT_{s}Q\left(  m_{s}\right)
,\;n=1,2,3,\cdots\\
b  &  \leq0:\;T_{-}Q\left(  m_{-}\right)  =nT_{s}Q\left(  m_{s}\right)
,\;n=1,2,3,\cdots
\end{align}
Note that in the limit $K\rightarrow0$ these solutions reduce to the solutions
for the flat case with $b\geq0$ and $b\leq0.$ We will not discuss them in any
more detail here since this $K<0$ case is similar to the previous discussion
for both $K>0$ and $K=0.$

\section{Summary and Outlook}

We have thoroughly analyzed a simple model of a scalar field interacting with
gravity in 3+1 dimensions. The model was derived from 2T-gravity in 4+2
dimensions as the \textquotedblleft3+1 dimensional conformal
shadow\textquotedblright\ \cite{2Tgravity}\cite{2TgravityGeometry}%
\cite{inflationBC} and can also be constructed$^{\text{\ref{Neil}}}$ in the
colliding branes scenario \cite{cyclicST} in 4+1 dimensions using the
worldbrane notions \cite{RandallSundrum} inspired by M-theory
\cite{HoravaWitten}.

An essential feature of the model in 3+1 dimensions is an underlying local
conformal symmetry (Weyl symmetry) exhibited in the action of
Eq.(\ref{initialS}). There is no fundamental gravitational constant in this
model, but instead there is a gauge dependent dynamical \textquotedblleft
gravitational parameter\textquotedblright\ which plays precisely the role of
the gravitational constant when the Weyl symmetry is gauge fixed to the
Einstein frame, thus agreeing with the standard form of Einstein's gravity and
its interactions with matter.

This raised the question of whether the dynamics could force the
\textquotedblleft gravitational parameter\textquotedblright\ $\left(  \phi
^{2}\left(  x^{\mu}\right)  -s^{2}\left(  x^{\mu}\right)  \right)  ^{-1}$ to
change sign in some patches of space-time where antigravity would emerge, and
whether the existence of such patches could have observational consequences in
our current universe in the context of cosmology or otherwise. This is the
question that motivated our investigation.

We found out that generically the gravitational parameter does change sign
dynamically, and that this change of sign is not a gauge artifact since the
gauge invariant quantity $\left(  1-s^{2}/\phi^{2}\right)  $ can be used to
monitor the sign change. So our model indicates that patches of antigravity
could exist, but we have also found that such patches cannot be reached from
our current universe without going through singularities, such as the big bang
or big crunch (perhaps others, such as black holes as well).

We have not yet answered what the physical implications of this phenomenon may
be for our current universe, but instead we have limited the current
investigation to finding and classifying those classical solutions in the
context of cosmology that are geodesically complete for all times. By
\textit{all} times we mean that one must go beyond the gauge dependent
definition of \textquotedblleft time\textquotedblright\ and instead seek
geodesically complete solutions in all possible choices of \textquotedblleft
time\textquotedblright, thus being able to connect information before and
after singularities. In this way one is not limited to only some patches while
declaring ignorance about other patches\footnote{Of course, we must expect
modifications of the classical equations due to quantum gravity especially
near singularities. We may need to wait a long time before one is sure about
what quantum gravity really is. Given this cloudiness of our knowledge at the
present time, we feel that the geodesically complete approach we are pursuing
here will still be relevant, and perhaps even provide guidance to clarify the
issues in future research.}.

We found that the conformal time $\tau$ is a good evolution parameter for our
purpose, so we analyzed the solutions for all values of $\tau$ from minus
infinity to plus infinity. In this way we learned that the gauge invariant
quantity $\left(  1-s^{2}\left(  \tau\right)  /\phi^{2}\left(  \tau\right)
\right)  $ oscillates back and forth between patches where it is positive and
negative (namely gravity/antigravity), and this information is carried
smoothly through the singularities. In fact we learned that the point in time
where there is a singularity in the Einstein frame (divergent scalar
curvature) does not look like a singularity at all in other convenient gauge
choices of the Weyl gauge symmetry.

This paper was focussed on finding and classifying the complete subset of all
classical cosmological solutions for which $\left(  1-s^{2}\left(
\tau\right)  /\phi^{2}\left(  \tau\right)  \right)  $ never changes sign for
all times. Thus, for the classical solutions exhibited in this paper the
universe remains always in the gravity patch, never shifting to antigravity.
The universe starts expanding with a big bang, but eventually it turns around
(at a finite or infinite size, depending on the sign of the parameter $b)$ and
begins to contract, leading to a big crunch. But this is followed with the
same periodic pattern of a big bang, turnaround, big crunch, again and again
indefinitely to the future as well as to the past.

There are such cyclic solutions in which the universe never contracts to zero
size, and bounces back after contracting to a finite size. So in these finite
bounce solutions the universe never hits a curvature singularity (as defined
in the Einstein frame). These solutions are possible for the closed universe.
The finite bounce solutions exist for a large range of the parameters $b,c,K$
that define the model, but one of the integration constants $E$ (which amounts
to energy initial conditions for the scalar fields $s$ or $\phi$) must lie in
a certain range defined by the parameters $b,c,K,$ while the other integration
constant $\tau_{0}$ is arbitrary.

There are also cyclic solutions in which the universe contracts to zero size
periodically, thus hitting the curvature singularity (in the Einstein frame)
at the big crunches/bangs. These cyclic solutions occur for the flat, open and
closed universes, but only if some initial conditions for the $\phi,s$ fields
are synchronized (the integration parameter $\tau_{0}$ set to $\tau_{0}=0$)
and a quantization condition is imposed on a combination of the parameters
$b,c,K$ and the integration parameter $E.$ Thus not every model is capable of
yielding geodesically complete cyclic solutions, as illustrated clearly in the
case of the flat universe.

Evidently the next stage of this research is to analyze what happens to these
solutions under perturbations. This is the topic of our next paper in
Ref.\cite{bcst1} where in addition to curvature $K,$ radiation and anisotropy
are included both at the classical and quantum (in the sense of the
Wheeler-deWitt equation) levels. According to our current understanding, very
similar geodesically complete solutions exist in the presence of these
perturbations. The question of the physics of antigravity and its effect on
our current era of cosmology is an interesting topic that we intend to pursue
as a natural evolution of the present discussion. We hope to report on these
details in the near future.

\begin{acknowledgments}
We thank Paul Steinhardt for stimulating conversations and encouragement. This
work was initiated at Princeton University and completed at the Perimeter
Institute. I. Bars and S.H.Chen would like to thank these institutions for
support and hospitality.
\end{acknowledgments}

\end{document}